\documentclass[journal,twoside,final,letterpaper,twocolumn]{IEEEtran}

\ifCLASSINFOpdf
\usepackage[pdftex]{graphicx}
\else

\fi

\usepackage{epstopdf}
\usepackage{epsfig}
\usepackage{float}
\usepackage{amssymb,amsbsy,subfigure,amsmath,amsfonts,multirow,color,url,bm,cite,cases,eqparbox}
\usepackage{algorithmic}
\usepackage{algorithm}
\usepackage[mathscr]{eucal}
\usepackage{enumitem}
\usepackage{amssymb}
\usepackage{soul}
\usepackage{xcolor}
\usepackage{bbding}

\newcommand{\tablesize}{\fontsize{6.5pt}{12pt}\selectfont}

\def\Xcal{\mathcal{X}}

\def\A{\mathbf{A}}
\def\B{\mathbf{B}}
\def\C{\mathbf{C}}

\def\K{\mathbf{K}}

\def\cited#1{{\iffalse #1 \fi}}
\def\1#1{\textcolor{red}{\textbf{#1}}}
\def\2#1{\textcolor{blue}{\textbf{#1}}}
\def\3#1{\textcolor{green}{\textbf{#1}}}

\begin{document}
	
	\title{SSUMamba: Spatial-Spectral Selective State Space Model for Hyperspectral Image Denoising}

	\author{Guanyiman Fu, Fengchao Xiong, \IEEEmembership{Member,~IEEE}, Jianfeng Lu, \IEEEmembership{Member,~IEEE} and Jun Zhou, \IEEEmembership{Senior Member,~IEEE}
\thanks {This work was supported in part by  the National Natural Science Foundation of China under Grant 62371237 and the Fundamental Research Funds for the Central Universities under Grant 30923010213. (Corresponding author: Fengchao Xiong.)}
\thanks{Guanyiman Fu, Fengchao Xiong  and Jianfeng Lu are with the School of Computer Science and Engineering, Nanjing University of Science and Technology, Nanjing 210094, China. }
\thanks{Jun Zhou is with the School of Information and Communication Technology, Griffith University, Nathan, Australia.}}

\maketitle
\begin{abstract}
	Denoising is a crucial preprocessing step for hyperspectral images (HSIs) due to noise arising from intra-imaging mechanisms and environmental factors. Long-range spatial-spectral correlation modeling is beneficial for HSI denoising but often comes with high computational complexity. Based on the state space model (SSM),  Mamba is known for its remarkable long-range dependency modeling capabilities and  computational efficiency. Building on this, we introduce a memory-efficient spatial-spectral UMamba (SSUMamba) for HSI denoising, with the spatial-spectral continuous scan (SSCS) Mamba being the core component.  SSCS Mamba alternates the row, column, and band in six different orders to generate the sequence and uses the bidirectional SSM to exploit long-range spatial-spectral dependencies. In each order, the images are rearranged between adjacent scans to ensure spatial-spectral continuity. Additionally, 3D convolutions are embedded into the SSCS Mamba to enhance local spatial-spectral modeling.   Experiments demonstrate that SSUMamba achieves superior denoising results with lower memory consumption per batch compared to transformer-based methods. The source code is available at https://github.com/lronkitty/SSUMamba.
\end{abstract}

%

\section{Introduction}
Hyperspectral images (HSIs) capture information across the electromagnetic spectrum, providing the spectrum for each pixel in an image. This spectral information enables the identification of objects and materials by analyzing their unique spectral signatures. As a result, HSIs have been extensively used in material recognition~\cite{Thai2002}, object detection~\cite{Yang2023,Gao2023,Dong2023}, object tracking~\cite{Li2023a,Xiong2020,Li2024b}, change detection~\cite{Luo2023}, and environmental protection~\cite{Li2024}. Noise is an unavoidable issue for HSI interpretation, arising from various factors such as insufficient exposure time, mechanical vibrations of imaging platforms, atmospheric perturbations, stochastic errors in photon counting, and other intrinsic and extrinsic elements~\cite{li2022spatial, Fu2024}. Hence, denoising stands as a pivotal preprocessing stage, enhancing image quality and improving the practical utility of HSIs.

Recent years have witnessed a shift from traditional model-based methods to deep learning (DL)-based ones, thanks to their powerful representation ability. Earlier convolutional neural network (CNN)-based methods achieved local spatial-spectral correlation modeling using 3-D convolutions~\cite{liu20193,Dong2019,Wang2023b}. Due to the large amount of missing information in degraded images, local modeling cannot adequately exploit the characteristics of HSIs to meet the demand for higher performance in denoising. Thanks to their capability to model long-range spatial and spectral correlations via self-attention, transformer-based methods have demonstrated stronger performance than CNN-based methods~\cite{li2022spatial, Pang2022}. Nevertheless, due to the $\text{O}(n^2)$ complexity of the self-attention mechanism, these methods often incorporate multiple modules to explore spatial and spectral domains separately, aiming to reduce the length of tokens and avoid redundant information.


\begin{figure}
	\centering
	\subfigure[]{\label{fig:ab1}\includegraphics[width=0.48\linewidth]{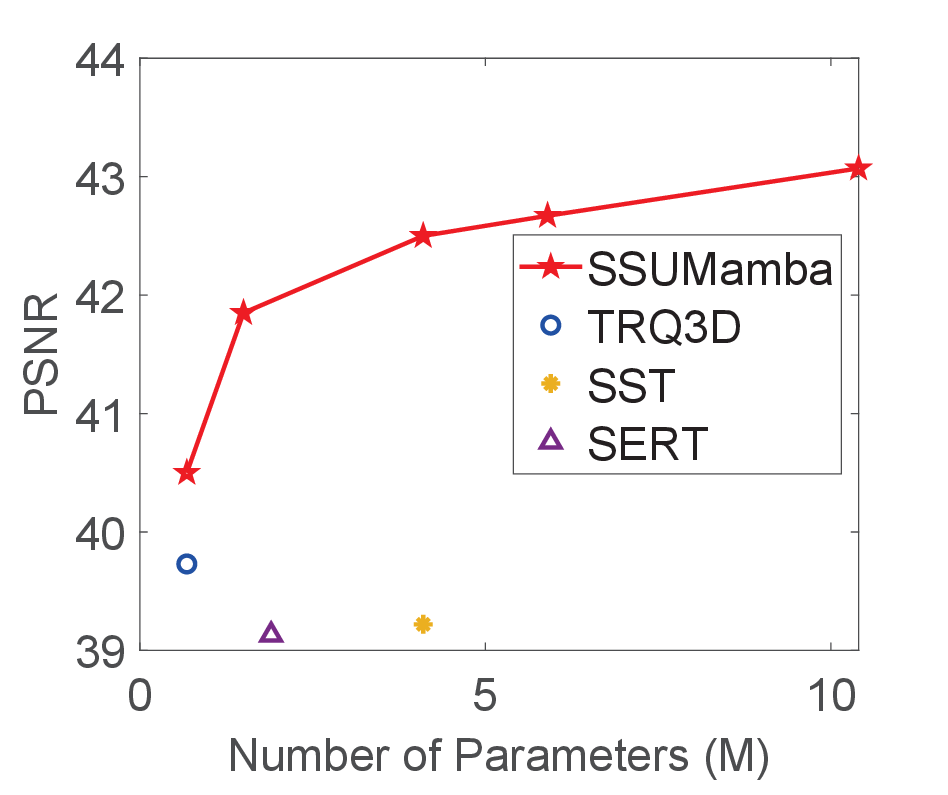}}
	\subfigure[]{\label{fig:ab2}\includegraphics[width=0.50\linewidth]{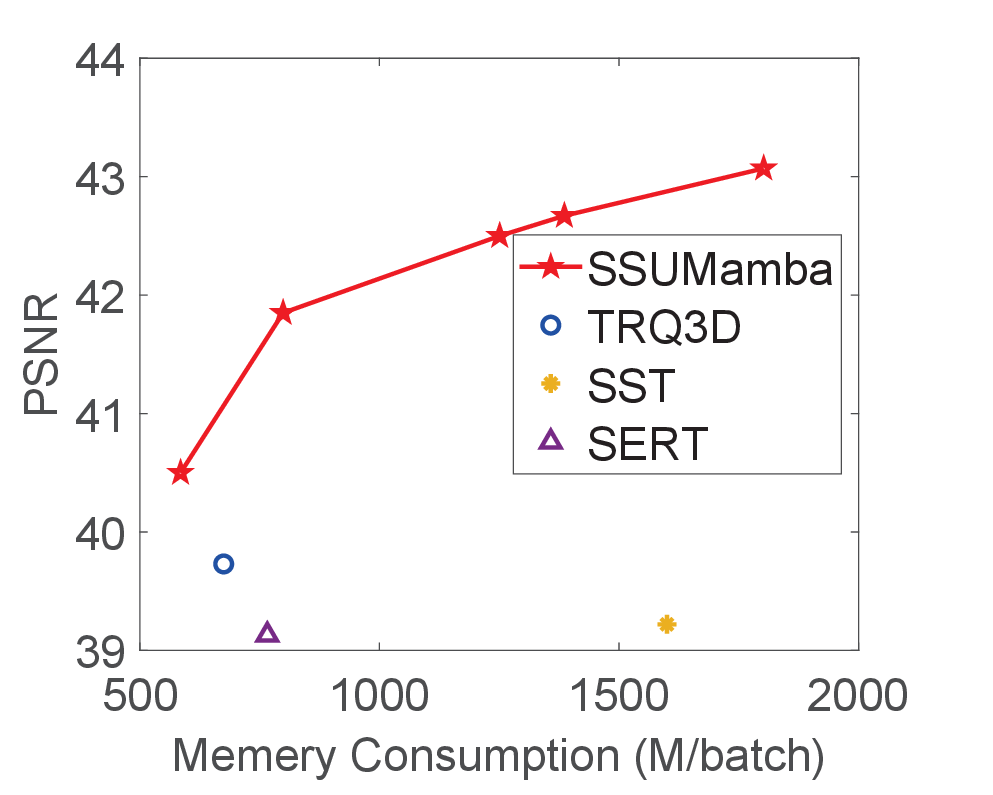}}
	\caption{Comparison of our SSUMamba and transformer-based TRQ3D, SST, and SERT  on the ICVL testing set.  The results indicate that SSUMamba outperforms transformer-based methods in peak signal-to-noise ratio (PSNR) while utilizing fewer parameters and memory consumption per batch.} \label{fig:ab}
\end{figure}

The state space model (SSM)\cite{gu2022efficiently,gu2023mamba}, rooted in control systems, has garnered attention for its linear complexity in handling long sequences. Recently, Mamba~\cite{gu2023mamba}, an optimized selective SSM, has demonstrated notable advantages in computer vision tasks~\cite{liu2024vmamba, li2024videomamba} due to its linear complexity, which offers promising benefits for long-range modeling in images. Unlike convolutional or self-attention operations, Mamba requires a specific ordering of the multidimensional data into 1D sequences, \emph{i.e.}, a scan scheme. An inappropriate scan can result in spatially close pixels being found at very distant locations in the flattened sequences, leading to the problem of local pixel forgetting. Given the non-sequential nature of images, bidirectional scanning~\cite{zhu2024vision} has been introduced to replace the unidirectional scan scheme, making Mamba more suitable for image processing. Zigzag scanning~\cite{hu2024zigma} further improves Mamba by incorporating spatially continuous scanning, ensuring data continuity during the ordering process. In the case of  HSIs, each pixel's spectral value depends not only on its neighboring pixels in the spatial domain but also on the corresponding pixels across different spectral bands. The unique properties of  HSI modeling pose significant challenges for modeling, making the application of Mamba for HSI denoising a relatively underexplored area. Liu~\emph{et al.}~\cite{liu2024hsidmamba} introduced an SSM-based method to exploit bidirectional long-range spatial dependencies in HSIs, but it overlooks long-range spectral dependencies. Therefore, applying Mamba to HSI denoising to capture long-range spatial-spectral correlations in 3-D HSIs remains an open question.

In this paper, we introduce a spatial-spectral U-Mamba (SSUMamba) for HSI denoising. SSUMamba features a U-shaped architecture and leverages the spatial-spectral continuous scan scheme (SSCS) to exploit long-range spatial-spectral dependencies. Building upon the zigzag scan, SSCS employs six bidirectional scanning schemes to generate multiple spatial-spectral continuous sequences for the subsequent state space model, thereby effectively capturing long-range spatial-spectral correlations. These schemes include row-column-band, row-band-column, column-row-band, column-band-row, band-row-column, and band-column-row orders. Based on the SSCS strategy, the spatial-spectral Mamba block further incorporates residual 3-D convolutional blocks to model local spatial-spectral dependencies, enhancing local texture exploration. Due to the linear complexity and SSCS in modeling long-range spatial-spectral dependencies, our SSUMamba method demonstrates superior denoising potential compared to CNN-based methods, which are flawed in modeling nonlocal information, and transformer-based methods, which split the HSI into windows or use multiple modules to separately capture spatial and spectral correlations.
Experimental results demonstrate the superior performance of our SSUMamba compared to other methods. To the best of our knowledge, this is the first time that the spatial-spectral Mamba has been introduced for the HSI denoising task.  In summary, the contributions of this work are as follows:

%
%

\begin{itemize}

	
	\item We introduce a spatial-spectral continuous scan (SSCS) Mamba block, which incorporates an SSCS strategy to generate continuous sequences, exploiting long-range spatial-spectral dependencies. Additionally, residual blocks are included to further model local spatial-spectral correlations and enhance texture preservation.
	
	\item Experiments demonstrate that SSUMamba based on the SSCS Mamba block achieves remarkable denoising capacity with fewer parameters and lower memory consumption per batch compared to transformer-based methods, as shown in Fig.~\ref{fig:ab}.

	\end{itemize}

The remainder of this paper is organized as follows. Section~\ref{sec:related} reviews related works. Section~\ref{sec:method} introduces the details of the proposed SSUMamba. Section~\ref{sec:experiments} presents quantitative and visual experiments. Finally, Section~\ref{sec:conclusion} concludes this paper.

\section{Related Work}\label{sec:related}

In this section, we offer a thorough review of HSI denoising, covering both model-based, CNN-based, and transformer-based methods. We also introduce the SSM in computer vision.

\subsection{HSI Denoising Approaches}
HSI denoising approaches can be categorized into model-based and DL-based methods. Conventional model-based methods exploit sparse and low-rank representations to describe the physical properties of clean HSIs for denoising~\cite{Fu2023,Zhang2014a,ye2014multitask,xiong2019, Zha2023,Su2023,Cai2023}. For example, BM4D~\cite{maggioni2012nonlocal}, an extension of BM3D~\cite{Dabov2007}, utilized collective sparse representation to capture similarity among nonlocal spatial-spectral cubes. LLRT~\cite{Chang2017} incorporated an analysis-based hyper-Laplacian prior into the low-rank model, effectively restoring HSIs while preserving their spectral structure. NGMeet~\cite{He2020} used low-rank Tucker decomposition and matrix rank minimization to capture global spectral correlation and nonlocal self-similarity, respectively. FastHyDe~\cite{zhuang2018fast} captures spatial-spectral representations by exploiting their sparse, low-rank, and self-similarity characteristics in subspaces.  Moreover, some works exploit the low-rankness of HSIs in the transformed domain to better learn a compact representation of HSIs~\cite{Zhang2023c,Chen2022,Wang2023}. E-3DTV~\cite{peng2020} discerned global correlations and local differences across all bands by evaluating the sparsity of gradient maps in subspaces. Zhang~\emph{et al.}~\cite{Zhang2023a} introduced a low-TR-rank regularizer and sparse regularizer to characterize the global low-rankness and sparsity of the transformed HSI, respectively, for denoising. These model-based approaches typically involve iterative optimization and manual hyperparameter fine-tuning, making them computationally inefficient and less practical for HSI denoising.

Since CNNs have demonstrated their power in low-level image processing~\cite{zhang2017beyond,Du2022,Du_2022_CVPR}, DL has emerged as a critical approach in HSI denoising~\cite{Li2024a,Pan2023,yuan2019}. Techniques utilizing 3-D convolution and its variations~\cite{liu20193,Dong2019,Wei2021} have played a crucial role in capturing local spatial-spectral correlations for HSI denoising. Liu~\emph{et al.}~\cite{liu20193} used a 3-D atrous CNN to expand the receptive field for HSI denoising, effectively handling mixture noise types while preserving image details. QRNN3D~\cite{Wei2021} integrated a recurrent unit after 3-D convolutions to leverage local spatial-spectral and global spectral correlation (GSC). NSSNN~\cite{guanyiman2022} combined an attention mechanism with the recurrent unit to extract GSC and nonlocal spatial features, achieving promising denoising performance. Wang~\emph{et al.}~\cite{Wang2024} introduced a two-stage network that leverages neighbor spectral maintenance and context enhancement for single HSI restoration. However, due to the locality representation capability of CNNs, these methods often fail to capture sufficient nonlocal information, resulting in suboptimal exploitation of the 3-D nature of HSI in recovery.

Thanks to the capability of the attention mechanism to model long-range information, transformers have been introduced into image denoising~\cite{zhou2024efficient}. Restormer~\cite{Zamir_2022_CVPR} incorporated channel attention, and SwinIR~\cite{Liang_2021_ICCV} applied attention in shifted windows to reduce the   $\text{O}(n^2)$ complexity caused by the attention mechanism. For HSI denoising~\cite{Chen2022_hider,Wang2022a}, intuitively, the 3-D transformer~\cite{Liu_2022_CVPR} seems like a straightforward solution for HSI denoising. However, the large number of bands significantly increases computational complexity and memory consumption in the attention mechanism, limiting the feasibility of using 3-D transformers for HSI denoising. As a compromise, transformers are often employed to capture one type of long-range correlation. For instance, TRQ3D~\cite{Pang2022} incorporated a shifted windowing scheme to model nonlocal spatial relations while using RNNs to model spectral correlation. Similarly, SST~\cite{li2022spatial} utilized windowing spatial and spectral transformers to model spatial and spectral correlation, respectively. To achieve longer-range modeling with limited space complexity, methods using different kinds of windows rather than square windows have been proposed. For example, Translution-SNet~\cite{Wang2022} used a cross-shaped window to exploit spatial correlation by merging horizontal and vertical linear patches. SERT~\cite{Li_2023_CVPR} split the HSIs into multiple cross-rectangles by merging horizontal and vertical rectangular patches instead of linear patches, achieving more focus on the informative neighboring pixels. Additionally, guiding mechanisms~\cite{Cai_2022_CVPR,Lai_2023_ICCV} allow the transformer to be more attentive to efficient spatial and spectral representations, enabling transformer-based methods to achieve better performance with limited spatial complexity. However, these methods often result in suboptimal spatial-spectral modeling of HSIs due to the destruction of image structure. Moreover, HSIs feature spectrally non-i.i.d. noise. Without selective relationship construction, the presence of harmful information, particularly from heavily polluted bands, can deteriorate denoising performance.\\

\subsection{State Space Model}

The state space model (SSM)\cite{gu2022efficiently} maps sequence data to a state space to capture long-term dependencies. Mamba, an improved version of SSM\cite{gu2023mamba}, introduces learnable parameters in a selective scan mechanism to choose relevant information in a data-dependent manner. Due to its exceptional computational efficiency and capability for modeling long-range dependencies, Mamba has demonstrated unique advantages in both computer vision tasks~\cite{liu2024vmamba, zhu2024vision, chen2024video, ma2024u} and natural language processing (NLP)\cite{Lieber2024}. To address the difference between non-sequential vision data and unidirectional sequence information, Liu~\emph{et al.}\cite{liu2024vmamba} introduced a vision Mamba block with a 2-D selective scan scheme to model spatial information in cross directions. Considering spatial continuity in vision data, the zigzag scan\cite{hu2024zigma} was introduced to enhance spatial continuity by exploiting patches in a zigzag pattern. In the field of image restoration, Shi~\emph{et al.}\cite{shi2024vmambair} proposed an omni selective scan mechanism to scan image features, including the channel dimension to comprehensively model  image. Guo~\emph{et al.}\cite{guo2024mambairsimplebaselineimage} employed a residue state space block to enhance the local dependencies and reduce channel redundancy. However, current methods based on Mamba are insufficient for HSI denoising, where dependencies exist in both spatial and spectral context pixels. In view of this, we apply the SSM for HSI denoising by considering spatial-spectral continuity in sequence generation to model long-range spatial-spectral correlations.


\section{Method}\label{sec:method}

In this section, we first introduce the preliminaries of SSM and Mamba. Then, we provide a detailed description of the proposed SSUMamba, including the overall architecture, the SSCS Mamba block, and the design details.

\subsection{Preliminary: State Space Models}

State space models (SSMs) conceptualize continuous systems that map a one-dimensional input sequence $x(t) \in \mathbb{R}^{M_1}$ to an output response $y(t) \in \mathbb{R}^{M_2}$. Mathematically, this is represented as:
\begin{equation}
    \begin{alignedat}{2}
    h^\prime(t) &= \A h(t) +& \B x(t) \\ \label{eq:ode}
    y(t) &= \C h(t)
    \end{alignedat}
\end{equation}
where $h(t) \in \mathbb{R}^N$ is the hidden state. Specifically, the hidden state is updated by the input $x(t)$ and the previous hidden state, determined by the evolution parameter $\A \in \mathbb{R}^{N \times N}$ and  projection parameter $\B \in \mathbb{R}^{N \times M_1}$.  The parameter $\A$ stores the historical information and determines the influence of the previous hidden state on the current hidden state. $\B$  determines how much the input $x(t)$ affects the hidden state.   The output $y(t)$ is generated by the hidden state $h(t)$ and the input $x(t)$, determined by $\C \in \mathbb{R}^{M_2 \times N}$, which describes how the hidden state is transformed into output.

In the case of a single-input single-output system, where $M_1 = M_2 = 1$, Eq.~(\ref{eq:ode}) can be discretized into $\text{SSM}(\bar{\A},\bar{\B}, \C)(x)$  as follows:

\begin{figure}[!t]
	\centering
	\includegraphics[width=1\linewidth]{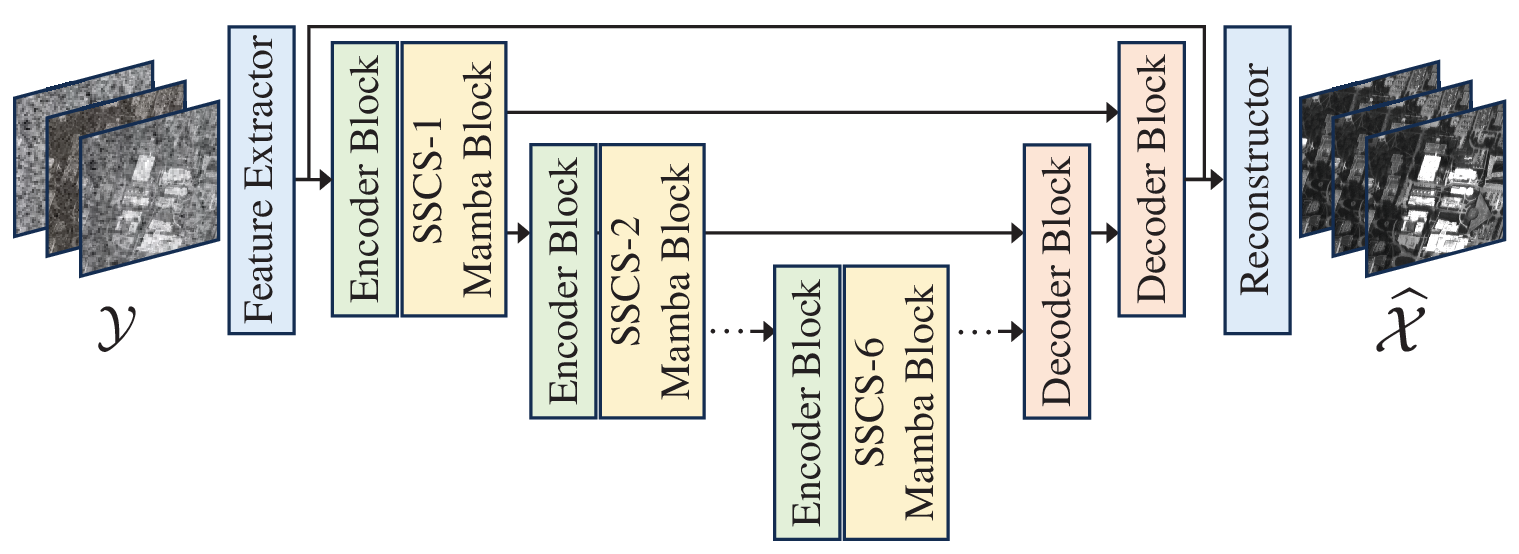}
	\caption{SSUMamba uses the encoder-decoder framework, incorporating a feature extractor, encoder blocks, SSCS Mamba blocks, decoder blocks with skip connections, and a reconstructor.}\label{fig:architecture}
  \end{figure}

  \begin{figure*}[ht]
	\centering
	\includegraphics[width=1\linewidth]{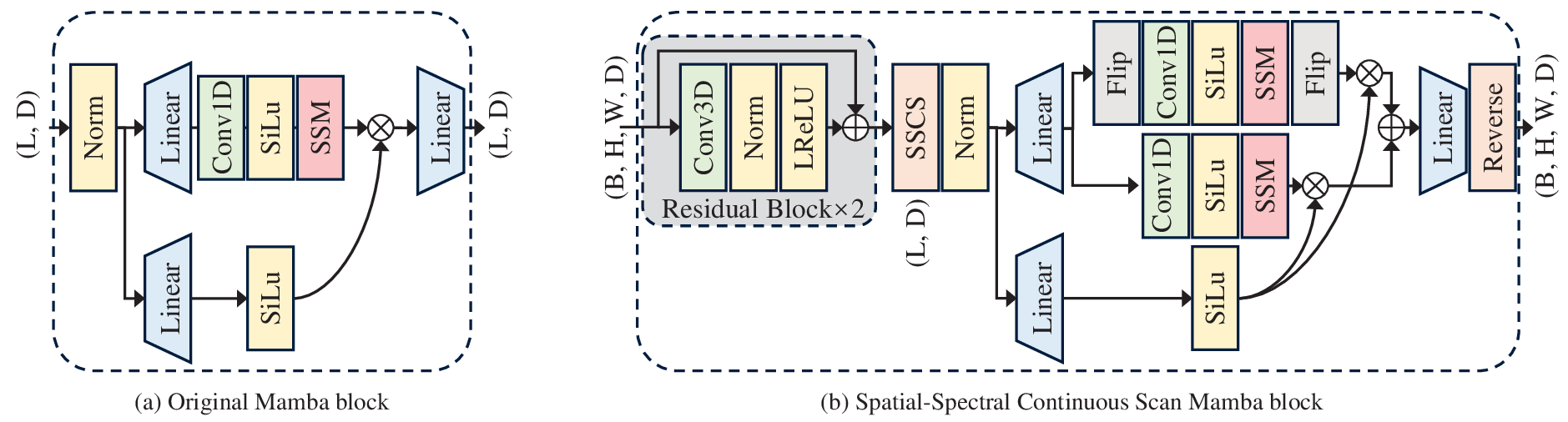}
	\caption{The original Mamba block and the introduced SSCS Mamba block. SSCS Mamba block contains residual blocks and a bidirectional SSM layer with SSCS.}\label{fig:mamba}
  \end{figure*}

\begin{equation}
	\begin{alignedat}{2}
	h_t &= \bar{\A}h_{t-1} + \bar{\B}x_t \\ \label{eq:zoh}
	y_t &= \bar{\C} h_t
	\end{alignedat}
\end{equation}
where $\bar{\A} \in \mathbb{R}^{N\times N}$, $\bar{\B} \in \mathbb{R}^{N\times 1}$  and $\bar{\C} \in \mathbb{R}^{1\times N}$ are the discrete parameters.  $\{\bar{\A}, \bar{\B} \}$ connects $\{\A, \B \}$ as follows:
\begin{equation}
	\begin{alignedat}{2}
	\bar{\A} &= e^{\Delta \A} \\ \label{eq:discretePara}
	\bar{\B} &= (\Delta \A)^{-1}(e^{\Delta \A}-\mathbf{I}) \cdot \Delta \B
	\end{alignedat}
\end{equation}
where $\Delta$ is the sampling step for discretization. To further accelerate the computation, the $\text{SSM}(\bar{\A},\bar{\B}, \C)(x)$ can be unrolled into a convolutional form as follows:
\begin{equation}
\begin{aligned}
\bar{\K} =& (\bar\C\bar{\B},\bar\C\bar{\A}\bar{\B},\bar\C\bar{\A}^2\bar{\B},\cdots,\bar\C\bar{\A}^{N-1}\bar{\B}) \\
y =& x\ast \bar{\K}
\end{aligned}
\end{equation}
where $\bar{\K} \in \mathbb{R}^{L}$ is a structured convolutional kernel, and $\ast$ denotes the convolution operation.

In Eq.~(\ref{eq:zoh}), parameters are fixed for all time steps, failing to account for varying importance at different steps. In practice, a context-aware model that selectively integrates necessary information into a sequential state is often desired. Mamba adopts a selective scan mechanism by making  $\B$, $\C$, and $\Delta$   functions of the inputs:

\begin{equation}
	\begin{alignedat}{2}
		&\bar{\B} = s_B(x)\\
		&\bar{\C} = s_C(x)\\
		&\Delta = \tau_{\Delta}(Parameter+s_\Delta(x))\\
	\end{alignedat}
\end{equation}
Such a manner enhances the model's ability to dynamically adjust its parameters based on the input, improving its performance in various contexts. In this paper, we apply Mamba to model the long-range spatial-spectral dependencies for HSI denoising.

\subsection{Overview Architecture}
Let $\mathcal{X}$ be a clean  HSI with $H \times W$ pixels and $B$ bands. When subjected to additive noise $\mathcal{E}$, such as Gaussian noise, impulse noise, strips, and deadlines, the observed HSI $\mathcal{Y}$ can be mathematically represented as:

\begin{equation}
\mathcal{Y} = \mathcal{X} + \mathcal{E}
\end{equation}
The proposed spatial-spectral U-Mamba (SSUMamba) is designed to learn the mapping from $\mathcal{Y}$ to $\mathcal{X}$ for denoising. The loss function of SSUMamba is defined as minimizing the $L_2$ distance between the predicted clean HSI, denoted as $\widehat{\mathcal{X}}$, and the ground truth clean HSI, denoted as $\mathcal{X}$:
\begin{equation}
	\mathcal{L}=\frac{1}{N}\sum_{i=1}^{N}\|\widehat{\Xcal}_i-\Xcal_i\|_F^2
 \end{equation}
 Here, $N$ represents the batch size within each iteration.

As illustrated in Fig.~\ref{fig:architecture}, SSUMamba  is constructed as follows:
\begin{equation}
	\begin{aligned}
		\mathcal{F}_0 &= \text{FeatureExtractor}(\mathcal{Y}) \\
		\mathcal{F}_i &= \text{SSCS-}i\text{-Mamba}(\text{Encoder}(\mathcal{F}_{i-1})), ~i=1,2,\cdots,L \\
		\mathcal{F}_i &= \text{Decoder}(\mathcal{F}_{i-1}+\mathcal{F}_{2L-i}), ~~~~~~i=L+1,\cdots,2L \\
		\widehat{\mathcal{X}} &= \text{Reconstructor}(\mathcal{F}_{2L}+\mathcal{F}_0)
	\end{aligned}
	\end{equation}
 Here, the  feature extractor is first utilized to extract an initial spatial-spectral feature $\mathcal{F}_0 \in \mathbb{R}^{B \times H \times W \times D}$ from HSI $\mathcal{Y}$, where $D$ represents the number of channels. The encoder blocks employ downsampling and 3-D convolutions to extract multi-scale features from the initial feature. The SSCS Mamba blocks, consisting of six schemes from SSCS-1 to SSCS-6, are employed to model local texture and long-range spatial-spectral dependencies. Then, the decoder blocks use upsampling and 3-D convolutions with residual connections to reconstruct the features. Finally, the reconstructor generates the denoised HSI  $\widehat{\mathcal{X}} \in \mathbb{R}^{B \times H \times W}$  from the decoded features using 3-D convolutions with a residual connection to the initial feature.

 \begin{figure}[ht]
	\centering
	\includegraphics[width=1\linewidth]{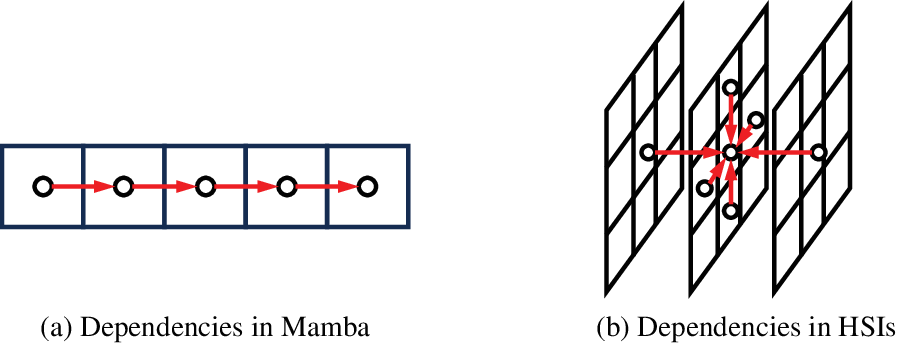}
	\caption{Illustration of the dependencies in Mamba and spatial-spectral dependencies in HSIs. The red arrows indicate the dependencies propagated.}\label{fig:dependencies}
  \end{figure}

\begin{figure*}[ht]
	\centering
	\includegraphics[width=1\linewidth]{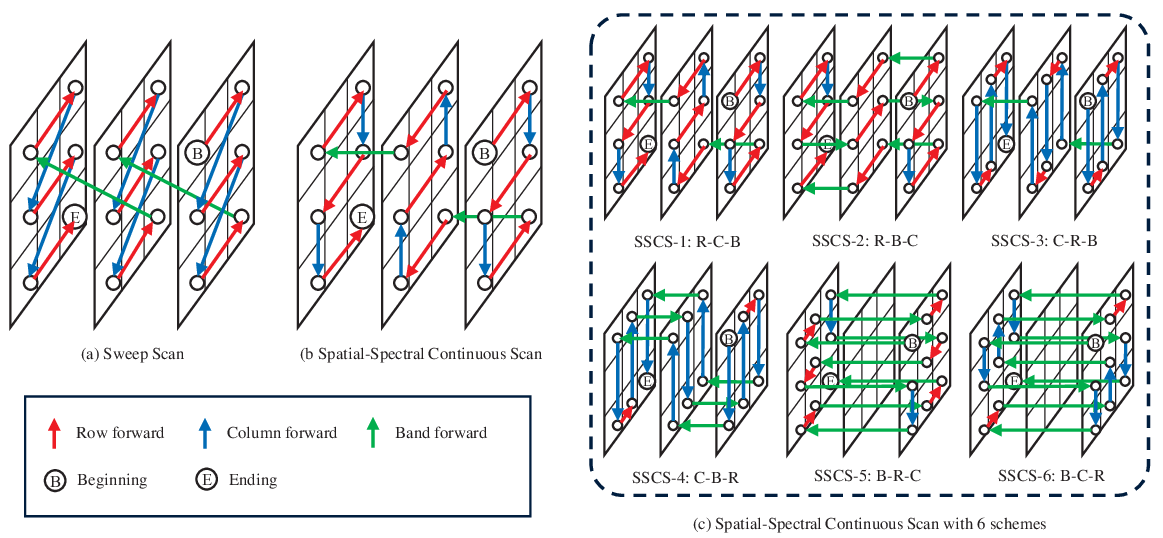}
	\caption{Illustration of our SSCS scheme.  (a) The sweep scan.  (b) A typical example of SSCS. (c) SSCS with  six schemes.}\label{fig:SSCS}
  \end{figure*}

\subsection{Spatial-Spectral Continuous Scan Mamba Block}

The Mamba block, illustrated in Fig.~\ref{fig:mamba}(a),   is renowned for its compelling features, such as achieving global receptive fields, dynamic weights, and linear computational complexity. However, it presents significant challenges when applied to 3D hyperspectral data, which exhibits spatial-spectral correlation. Following Eq.~(\ref{eq:zoh}), the current output of Mamba is generated by the hidden state from the last token and the current input token, resulting in dependencies being propagated in a unidirectional manner, as shown in  Fig.~\ref{fig:dependencies}(a). In HSIs, neighboring tokens often share similarities, reflecting strong dependencies in spatial and spectral domains, as shown in Fig.~\ref{fig:dependencies}(b). Unidirectional processing can cause spatially or spectrally close tokens to be found at very distant locations in the 1-D sequence, necessitating a scan order tailored to handle these dependencies effectively. Moreover, most Mamba models originate from NLP tasks characterized by a sequential nature and employ a sweep scan scheme where tokens follow the computer hierarchy, typically in a row-column-major order. As shown in Fig.~\ref{fig:SSCS}(a), the naive adaptation of such schemes for HSIs involves a row-column-band-major order. However, this approach neglects the spatial continuity of adjacent columns and bands \cite{hu2024zigma, liu2024vmamba, zhu2024vision}. Additionally, maintaining local spatial-spectral correlation is crucial for preserving image texture.

To address these challenges, we introduce the spatial-spectral continuous scan (SSCS) Mamba block for HSI modeling. As shown in Fig.~\ref{fig:mamba}(b), the SSCS Mamba block integrates residual blocks and a bidirectional SSM with SSCS to effectively capture both local and global spatial-spectral correlations and continuity.

\noindent \textbf{Residual Block:} The residual block enhances the local texture exploration by maintaining spatial-spectral correlation within a localized context. This block    includes a 3-D convolution, an Instance Normalization, and a Leaky ReLU (LReLU) and is applied to the input feature $\mathcal{F}_{\text{in}}$ via:
\begin{equation}
	\begin{aligned}
		\mathcal{F}_{\text{res}} = \text{LReLU}(\text{Norm}(\text{Conv3D}(\mathcal{F}_{\text{in}})))+\mathcal{F}_{\text{in}}
	\end{aligned}
\end{equation}

\noindent \textbf{Spatial-Spectral Continuous Scan:}  As shown  in Fig.~\ref{fig:SSCS}(b),   the spatial-spectral continuous scan (SSCS) rearranges HSIs in a zigzag scheme to account for continuity in both spatial and spectral domains during sequence generation. Taking the row-column-band order as an example, SSCS scans the input feature $\mathcal{F} \in \mathbb{R}^{B \times H \times W \times D}$ into a sequence $S \in \mathbb{R}^{L \times D}$, where $L=W \times H \times B$. The process involves scanning each band in a row-column order, which includes row-forward scans and column-forward scans. Between bands, a band-forward scan is employed. The row-forward and column-forward scans are connected in tail-to-tail or head-to-head connections, unlike the tail-to-head connections used in the sweep scan shown in Fig.~\ref{fig:SSCS}(a).   This is also the case with the column-forward and band-forward scans. This approach ensures that spatial-spectral continuity is maintained in the resulting sequences.

In addition, each pixel in the HSI relies on dependencies from surrounding pixels in different directions, including row, column, and band. To effectively model long-range spatial-spectral dependencies, we alternate these three directions and create six scan schemes: R-C-B, R-B-C, C-R-B, C-B-R, B-R-C, and B-C-R, where B, C, and R represent the band, column, and row, respectively. These six continuous scan schemes are illustrated in Fig.~\ref{fig:SSCS}(c). Each SSCS scheme generally requires different SSM parameters, and scaling up the number of schemes results in a significant increase in the number of parameters. To manage this complexity, we distribute the six schemes across six layers of the SSUMamba and alternate the sequence order between each layer, as depicted in Fig.~\ref{fig:architecture}, from SSCS-1 to SSCS-6.

\noindent\textbf{Bidirectional SSM:} To further enhance bidirectional dependencies, a bidirectional SSM layer flips the generated sequence to facilitate backward scanning, effectively exploiting long-range spatial-spectral dependencies in 12 directions. Formally, the bidirectional SSM with SSCS can be expressed as follows:

\begin{equation}
	\begin{aligned}
		S =& \text{SSCS}(\mathcal{F})\\
		\mathcal{G} =& \text{SiLu}(\text{Linear}(\text{Norm}(S)))\\
		S_L =& \text{Linear}(\text{Norm}(S) \\
		\mathcal{F}_f =& \text{SSM}(\text{SiLu}(\text{Conv1D}(S_L))) \\
		\mathcal{F}_b =& \text{Flip}(\text{SSM}(\text{SiLu}(\text{Conv1D}(\text{Flip}(\mathcal{S}_L))))) \\
		\mathcal{F}_{\text{out}} =& \text{Reverse}(\text{Linear}(\mathcal{F}_f \odot \mathcal{G} \  + \mathcal{F}_b \odot \mathcal{G}))
	\end{aligned}
\end{equation}
where selective mechanism, based on linear layers denoted as $\text{Linear}(\cdot)$, along with the gate $\mathcal{G}$ in the SSM, enables the model to selectively propagate or ignore dependencies. This mechanism assists the SSM in focusing on valuable information within the degraded sequence during HSI denoising. Finally, the output feature $\mathcal{F}_{\text{out}}$ is reshaped from the bidirectional SSM layer using a reverse operation of the SSCS.

\begin{table*}[htbp]
	\centering
	\caption{Comparison of Different Methods on 50 Testing HSIs from ICVL Dataset. The Top Three Values Are Marked as \1{Red}, \2{Blue}, and \3{Green}.}\label{tab:icvl}
	\resizebox{\linewidth}{!}{\tablesize{
	\begin{tabular}{c|c|c|c|c|c|c|c|c|c|c|c|c|c|c|c}
	   \hline
	&&&\multicolumn{6}{c|}{\textbf{Model-based methods}}&\multicolumn{7}{c}{\textbf{Deep learning-based methods}}\\
	\hline
	\multirow{2}*{\makebox[0.02\textwidth][c]{$\sigma$}}&\multirow{2}*{\makebox[0.02\textwidth][c]{Index}}&\multirow{2}*{\makebox[0.035\textwidth][c]{Noisy}}&\multirow{1}*{\makebox[0.035\textwidth][c]{BM4D}}&\multirow{1}*{\makebox[0.035\textwidth][c]{MTSNMF}}&\multirow{1}*{\makebox[0.035\textwidth][c]{NGMeet}}&\multirow{1}*{\makebox[0.035\textwidth][c]{FastHyDe}}&\multirow{1}*{\makebox[0.035\textwidth][c]{LRTF$L_0$}}&\multirow{1}*{\makebox[0.035\textwidth][c]{E-3DTV}}&\multirow{1}*{\makebox[0.035\textwidth][c]{T3SC}}&\multirow{1}*{\makebox[0.035\textwidth][c]{MAC-Net}}&\multirow{1}*{\makebox[0.035\textwidth][c]{NSSNN}}&\multirow{1}*{\makebox[0.035\textwidth][c]{TRQ3D}}&\multirow{1}*{\makebox[0.035\textwidth][c]{SST}}&\multirow{1}*{\makebox[0.035\textwidth][c]{SERT}}&\makebox[0.045\textwidth][c]{\textbf{SSUMamba}}\\
	&&&\cite{maggioni2012nonlocal}&\cite{ye2014multitask}&\cite{He2020}& \cite{zhuang2018fast}&\cite{xiong2019}& \cite{peng2020}& \cite{Bodrito2021}& \cite{xiong2021mac}& \cite{guanyiman2022}& \cite{Pang2022}& \cite{li2022spatial} & \cite{Li_2023_CVPR} & (ours)\\
	\hline
	\multirow{3}*{\makebox[0.02\textwidth][c]{[0,15]}}
	& \makebox[0.02\textwidth][c]{PSNR$\uparrow$}& 33.18 & 44.39 & 45.39 & 39.63 & 48.08 &43.41& 46.05   & 49.68  & 48.21   &49.83 & 46.43  & \2{50.87}  &\3{50.18} & \1{51.34}  \\
	& \makebox[0.02\textwidth][c]{SSIM$\uparrow$}& .6168  & .9683  & .9592    & .8612  & .9917 & .9315& .9811   & .9912 & .9915  &.9934& .9878 & \3{.9938}  &\1{.9977} & \2{.9946} \\
	& \makebox[0.02\textwidth][c]{SAM$\downarrow$} & .3368  & .0692  & .0845   & .2144  & .0404  & .0570	& .0560   & .0486 & .0387  &.0302 & .0437 & \3{.0298}  &\2{.0278} & \1{.0256}\\
	\hline 		
	\multirow{3}*{\makebox[0.02\textwidth][c]{[0,55]}}
	& \makebox[0.02\textwidth][c]{PSNR$\uparrow$}	& 21.72 & 37.63 & 38.02  & 31.53  & 42.86 &35.63 &	40.20   & 45.15  & 43.74 &46.27   & 44.64  & \2{46.39}  &\3{46.34} & \1{46.85}  \\
	& \makebox[0.02\textwidth][c]{SSIM$\uparrow$}	& .2339  & .9008  & .8586    & .6785    & .9800 & .8125& .9505   & .9810 & .9768&.9868  & .9840 & \3{.9872}  &\1{.9951} & \2{.9882} \\
	& \makebox[0.02\textwidth][c]{SAM$\downarrow$} 	& .7012  & .1397  & .234     & .4787    & .0630  & .1914	& .0993   & .0652 & .0582 &.0393  & .0487 & \3{.0457}  &\1{.0373} & \2{.0375}\\ 
	\hline
	\multirow{3}*{\makebox[0.02\textwidth][c]{[0,95]}}
	& \makebox[0.02\textwidth][c]{PSNR$\uparrow$}& 17.43 & 34.71 & 34.81 & 27.62 & 40.84 &32.83	&37.80   & 43.10  & 41.24 &44.42  & 43.54  & \2{44.83}  &\3{44.47} & \1{45.36}  \\
	& \makebox[0.02\textwidth][c]{SSIM$\uparrow$}& .1540   & .8402  & .7997   & .5363   & .9734  & .7482	& .9279   & .9734 & .9577 &.9809  & .9806 & \3{.9838}  &\1{.9929} & \2{.9853} \\
	& \makebox[0.02\textwidth][c]{SAM$\downarrow$} & .8893  & .1906  & .3266    & .6420     & .0771  & .3014	& .1317   & .0747 & .0841 &.0524  & .0523 & \3{.0513}  &\2{.0447} & \1{.0439} \\
	\hline 	
	\multirow{3}*{\makebox[0.02\textwidth][c]{Mixture}}
	& \makebox[0.02\textwidth][c]{PSNR$\uparrow$}& 13.21 & 23.36 & 27.55  & 23.61  & 27.58 &30.93	&34.90   & 34.09  & 28.44 &\2{40.54}  & \3{39.73}  & 39.22   &39.13 & \1{43.07}  \\
	& \makebox[0.02\textwidth][c]{SSIM$\uparrow$}& .0841  & .4275  & .6743   & .4448    & .7250 & .8378	& .9041   & .9052 & .7393 &.9560 & .9491 & \3{.9626}  &\2{.9679} & \1{.9726} \\
	& \makebox[0.02\textwidth][c]{SAM$\downarrow$} & .9124  & .5476  & .5326    & .6252    & .4534 & .3613	& .1468   & .2340 & .4154 &.1097 & \3{.0869} & \2{.0743}  &.0963 & \1{.0710}\\
	\hline
	\end{tabular}}}
	\end{table*}

\section{Experiments}\label{sec:experiments}
In this section, we present the denoising performance of the proposed SSUMamba through synthetic and real-world experiments. Additionally, we conduct further ablation experiments to analyze the performance of each module and investigate the impact of varying widths within the network.

\subsection{Experimental Settings}

\subsubsection{Training and Testing Setting}

We selected 100 hyperspectral images (HSIs) from the Interdisciplinary Computational Vision Laboratory (ICVL) HSI dataset for training, captured using a Specim PS Kappa DX4 hyperspectral camera. These HSIs have a resolution of $1392 \times 1300$ pixels and cover 31 spectral bands ranging from 400 to 700 nm. During training, we partitioned these images into patches of size $64 \times 64 \times 31$ and further augmented the training set through random flipping, cropping, and resizing.

The trained model was evaluated on synthetic and real-world datasets. Synthetic datasets included the ICVL testing set, Houston 2018 HSI, and Pavia City Center HSI. The ICVL testing set consists of 50 HSIs, each with a resolution of $512 \times 512 \times 31$. The Houston 2018 HSI, captured using the ITRES CASI hyperspectral imager, comprises $1202 \times 4172$ pixels and 48 bands ranging from 380 to 1050 nm. For our experiments, we excluded the first 2 bands due to noise and extracted a $512 \times 512$ cube from the center of the HSI. The Pavia City Center HSI, acquired using the Reflective Optics System Imaging Spectrometer (ROSIS) sensor, covers the range of 430 to 860 nm. Following the preprocessing steps described in~\cite{He2015}, we removed the first 22 noisy bands and added mixture noise to a $200 \times 200$ subimage with 80 bands.

Real-world datasets used for evaluation included the Gaofen-5 Wuhan HSI and the Earth Observing-1 HSI. The Gaofen-5 Wuhan HSI, acquired using the Advanced Hyperspectral Imager (AHSI), has a resolution of $300 \times 300$ pixels and 155 spectral bands. The Earth Observing-1 HSI, captured by the Hyperion sensor, comprises 242 bands with a resolution of $1000 \times 400$ pixels. Following the preprocessing steps described in~\cite{Zhang2014a}, we removed water absorption bands, resulting in a final HSI cropped to $400 \times 200$ pixels with 166 bands.

It's important to note that T3SC, TRQ3D, SST, and SERT may face challenges when handling HSIs with distinct bands compared to the training data. To address this, we partitioned the provided HSI into multiple sub-images, each containing 31 bands, for testing.

\subsubsection{Noise Pattern}

As indicated by~\cite{chen2018}, real-world noises in HSIs typically follow a non-independent and identical distribution (non-i.i.d.). Additionally, remote-sensing HSIs are often contaminated by a mixture of different noise types. Consequently, our experiments considered both non-i.i.d. Gaussian noise and mixture noise.

For non-i.i.d. Gaussian noise, we varied the standard deviation within the ranges of $\sigma \in$ [0, 15], [0, 55], and [0, 95]. The mixture noise comprised the following components: 1) non-i.i.d. Gaussian noise with $\sigma \in$ [0, 95]; 2) impulse noise affecting 1/3 of the bands, with intensities spanning from 10\% to 70\%; 3) strips on 5\% to 15\% of columns on 1/3 of the bands; and 4) deadlines on 5\% to 15\% of columns on 1/3 of the bands.

\begin{table*}[htbp]
	\caption{Comparison of Different Methods on Houston 2018 HSI. The Top Three Values Are Marked as \1{Red}, \2{Blue}, and \3{Green}.}\label{tab:houston}
	\centering
	\resizebox{\linewidth}{!}{\tablesize{
		\begin{tabular}{c|c|c|c|c|c|c|c|c|c|c|c|c|c|c}
			\hline
		 &&\multicolumn{6}{c|}{\textbf{Model-based methods}}&\multicolumn{7}{c}{\textbf{Deep learning-based methods}}\\
		 \hline
		 \multirow{2}*{\makebox[0.02\textwidth][c]{Index}}&\multirow{2}*{\makebox[0.035\textwidth][c]{Noisy}}&\multirow{1}*{\makebox[0.035\textwidth][c]{BM4D}}&\multirow{1}*{\makebox[0.035\textwidth][c]{MTSNMF}}&\multirow{1}*{\makebox[0.035\textwidth][c]{NGMeet}}&\multirow{1}*{\makebox[0.035\textwidth][c]{FastHyDe}}&\multirow{1}*{\makebox[0.035\textwidth][c]{LRTF$L_0$}}&\multirow{1}*{\makebox[0.035\textwidth][c]{E-3DTV}}&\multirow{1}*{\makebox[0.035\textwidth][c]{T3SC}}&\multirow{1}*{\makebox[0.035\textwidth][c]{MAC-Net}}&\multirow{1}*{\makebox[0.035\textwidth][c]{NSSNN}}&\multirow{1}*{\makebox[0.035\textwidth][c]{TRQ3D}}&\multirow{1}*{\makebox[0.035\textwidth][c]{SST}}&\multirow{1}*{\makebox[0.035\textwidth][c]{SERT}}&\makebox[0.045\textwidth][c]{\textbf{SSUMamba}}\\
		 &&\cite{maggioni2012nonlocal}&\cite{ye2014multitask}&\cite{He2020}& \cite{zhuang2018fast}&\cite{xiong2019}& \cite{peng2020}& \cite{Bodrito2021}& \cite{xiong2021mac}& \cite{guanyiman2022}& \cite{Pang2022}& \cite{li2022spatial} & \cite{Li_2023_CVPR} & (ours)\\
		 \hline
	PSNR$\uparrow$  & 11.72 & 22.76 & 25.86 & 22.36 & 27.07 & 28.75 & 30.64 & 29.84  &28.83&\2{33.77}& \3{32.55} & 31.07 &31.31& \1{34.74} \\
	SSIM$\uparrow$  & .0843 & .4762 & .6933 & .5169 & .7757 &.8038& .8570 & .8751  &.7963&\1{.9457}& .9194 & .9166 &\3{.9296}& \2{.9452} \\
	SAM$\downarrow$ & .9778 & .5168 & .4977 & .5728 & .4518 &.2221& .1323 & .1943  &.2356&\2{.1094}& \3{.1241} & .1390 &.1517& \1{.0993} \\
	\hline	
	\end{tabular}}}
\end{table*}

\begin{figure*}[htbp]
	\centering
	\subfigure[Clean]{\label{fig:houston_clean}\includegraphics[width=0.124\linewidth]{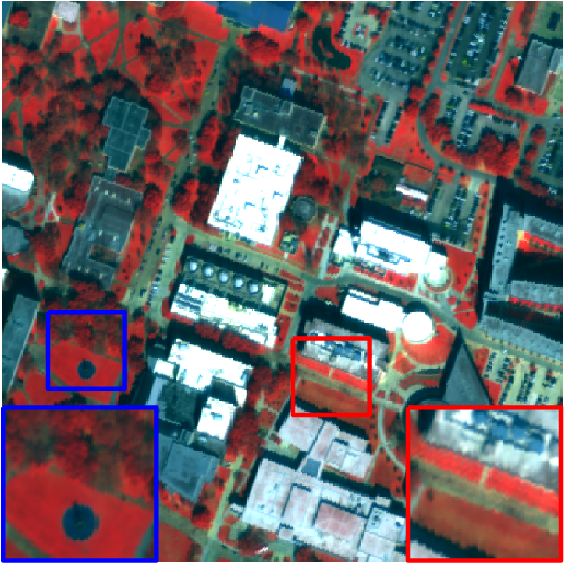}}
	\hspace{-2.1mm}
	\subfigure[Noisy]{\label{fig:houston_noisy}\includegraphics[width=0.124\linewidth]{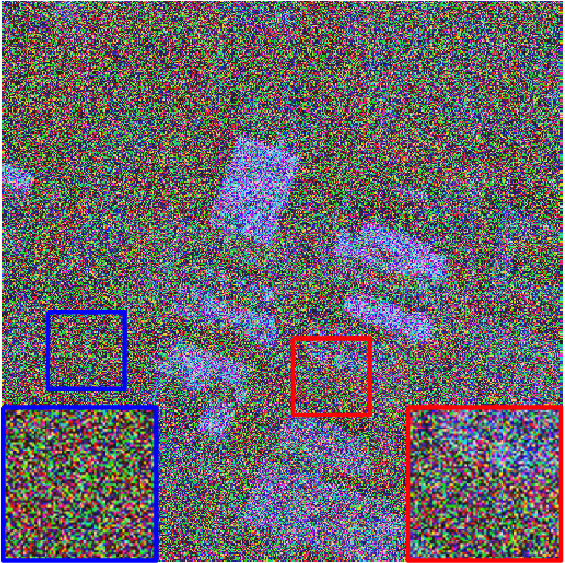}}
	\hspace{-2.1mm}
	\subfigure[BM4D \cite{maggioni2012nonlocal}]{\label{fig:houston_BM4D}\includegraphics[width=0.124\linewidth]{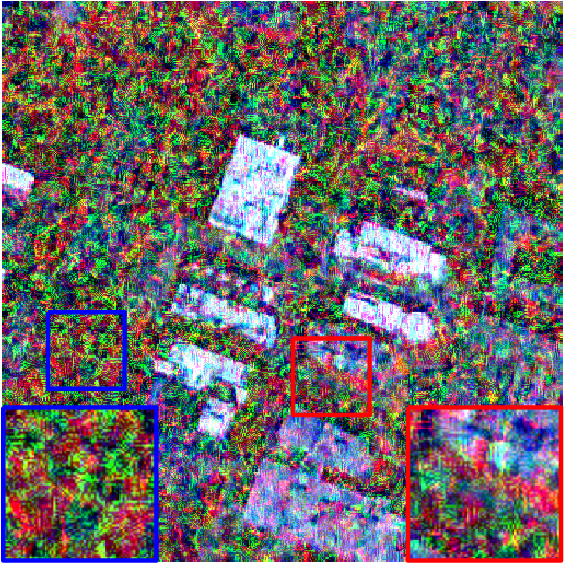}}
	\hspace{-2.1mm}
	\subfigure[MTSNMF \cite{ye2014multitask}]{\label{fig:houston_MTSNMF}\includegraphics[width=0.124\linewidth]{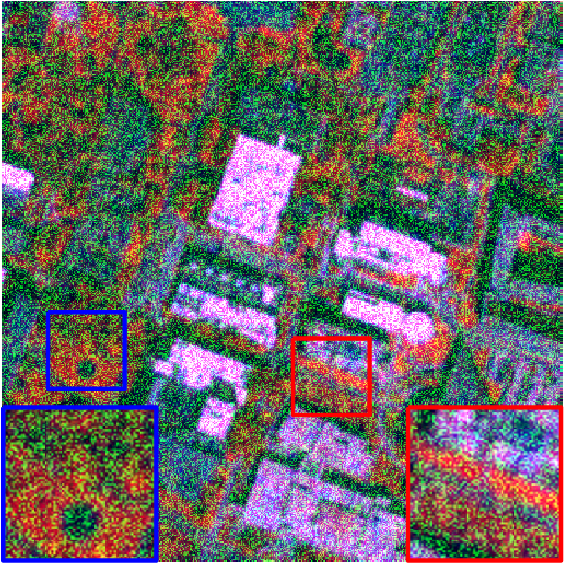}}
	\hspace{-2.1mm}
	\subfigure[NGMeet \cite{He2020}]{\label{fig:houston_NGMeet}\includegraphics[width=0.124\linewidth]{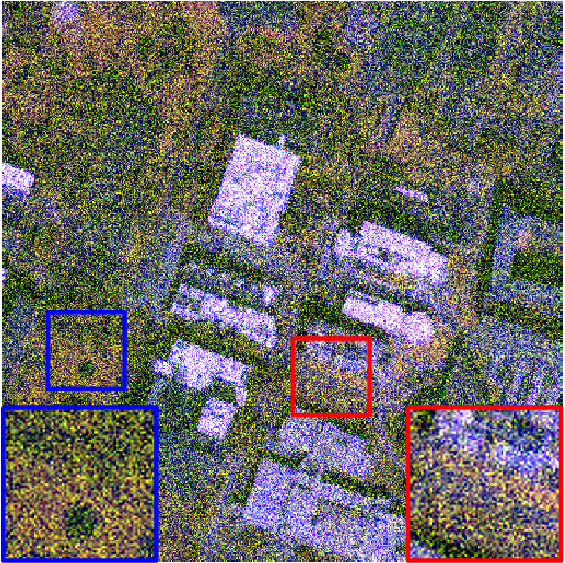}}
	\hspace{-2.1mm}
	\subfigure[FastHyDe \cite{zhuang2018fast}]{\label{fig:houston_FastHyDe}\includegraphics[width=0.124\linewidth]{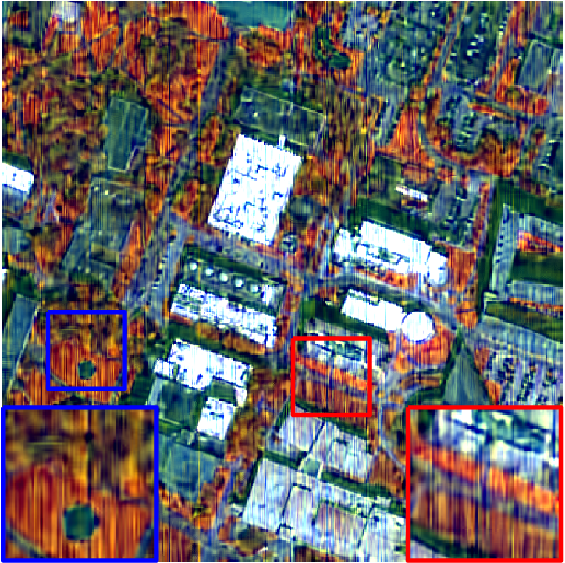}}
	\hspace{-2.1mm}
	\subfigure[LRTF$L_0$ \cite{xiong2019}]{\label{fig:houston_lrtfl0}\includegraphics[width=0.124\linewidth]{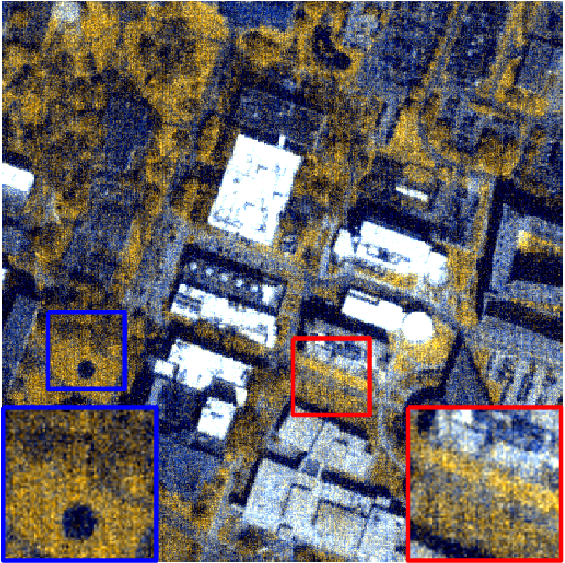}}
	\hspace{-2.1mm}
	\subfigure[E-3DTV \cite{peng2020}]{\label{fig:houston_e3dtv}\includegraphics[width=0.124\linewidth]{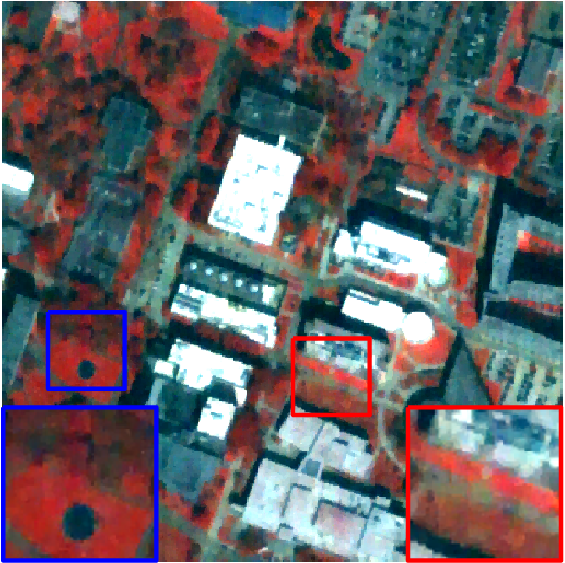}}\\
	\subfigure[T3SC \cite{Bodrito2021}]{\label{fig:houston_T3SC}\includegraphics[width=0.142\linewidth]{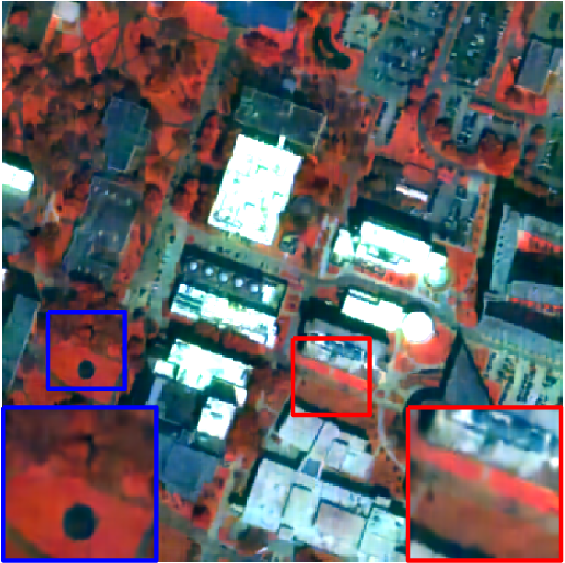}}
	\hspace{-2.1mm}
	\subfigure[MAC-Net \cite{xiong2021mac}]{\label{fig:houston_MAC-Net}\includegraphics[width=0.142\linewidth]{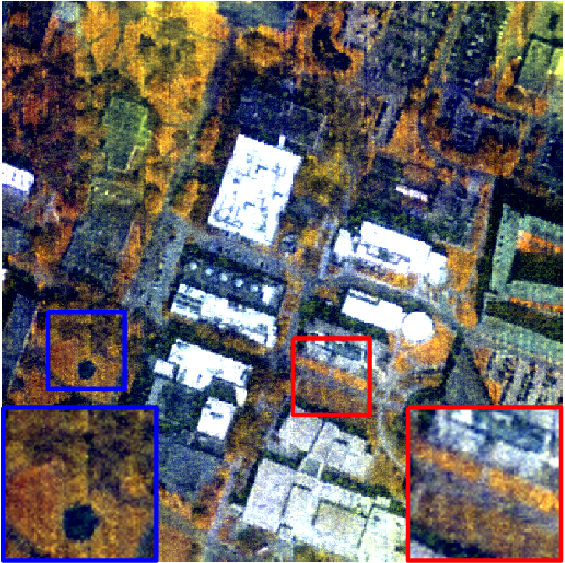}}
	\hspace{-2.1mm}
	\subfigure[NSSNN \cite{guanyiman2022}]{\label{fig:houston_NSSNN}\includegraphics[width=0.142\linewidth]{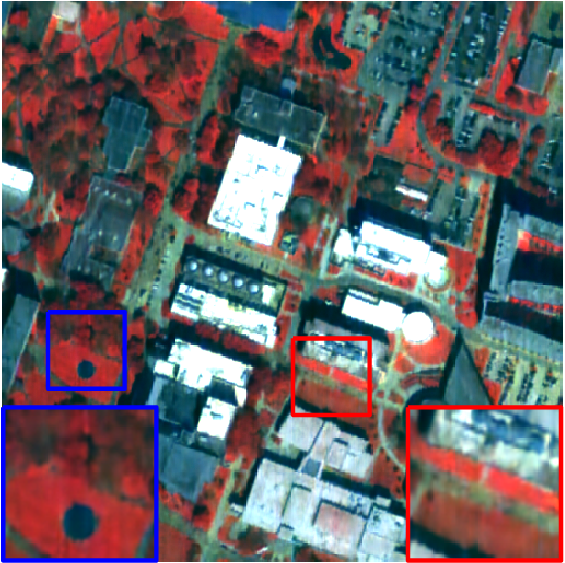}}
	\hspace{-2.1mm}
	\subfigure[TRQ3D \cite{Pang2022}]{\label{fig:houston_TRQ3D}\includegraphics[width=0.142\linewidth]{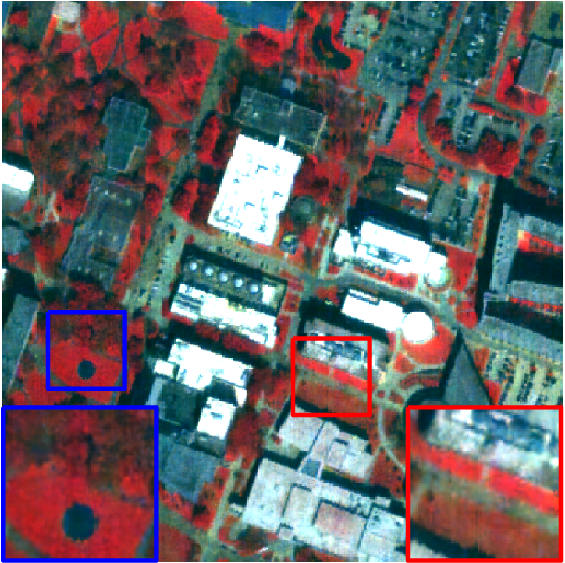}}
	\hspace{-2.1mm}
	\subfigure[SST \cite{li2022spatial}]{\label{fig:houston_SST}\includegraphics[width=0.142\linewidth]{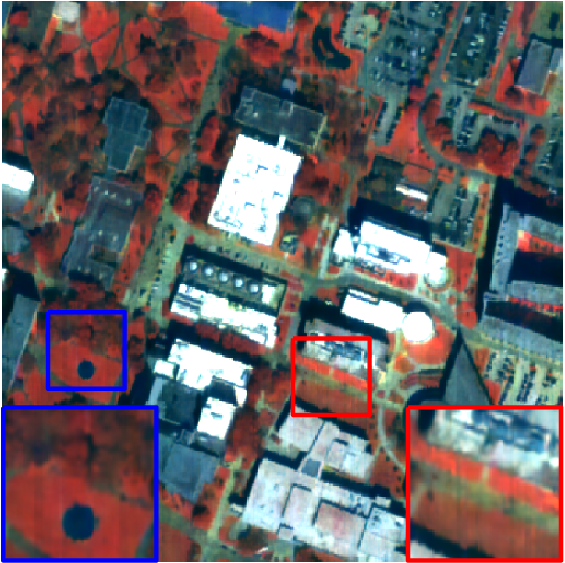}}
	\hspace{-2.1mm}
	\subfigure[SERT \cite{Li_2023_CVPR}]{\label{fig:houston_SERT}\includegraphics[width=0.142\linewidth]{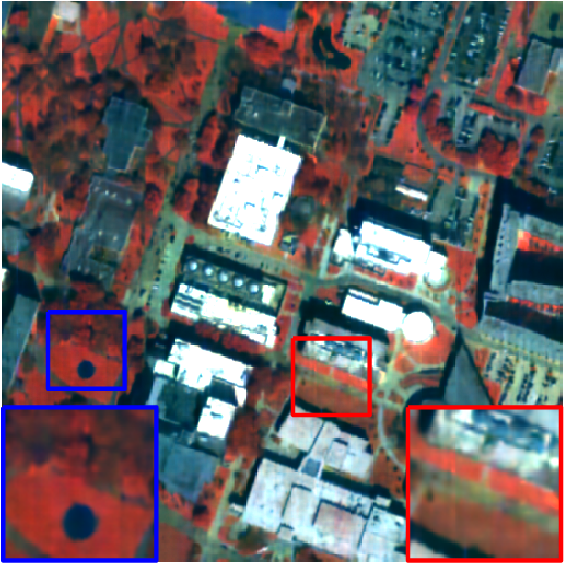}}
	\hspace{-2.1mm}
	\subfigure[\textbf{SSUMamba}]{\label{fig:houston_SSUMamba}\includegraphics[width=0.142\linewidth]{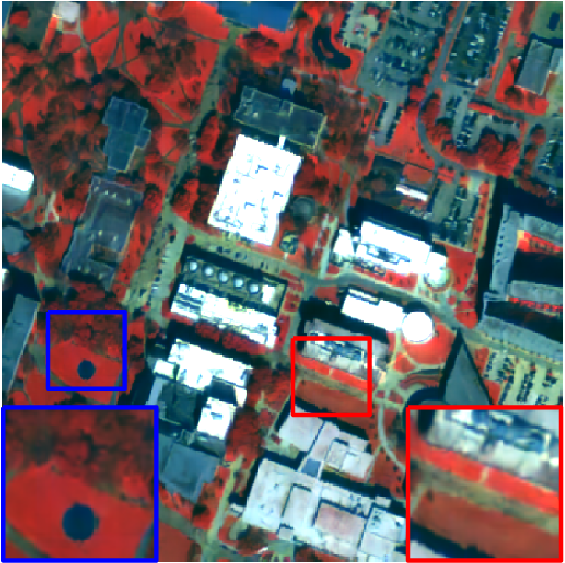}}
	   \caption{Denoising results on the Houston 2018 HSI with the mixture noise. The false-color images are generated by combining bands 35, 20, and 5.} \label{fig:houston_visual}
 \end{figure*}

\begin{figure}[ht]
	\centering
	\includegraphics[width=0.85\linewidth]{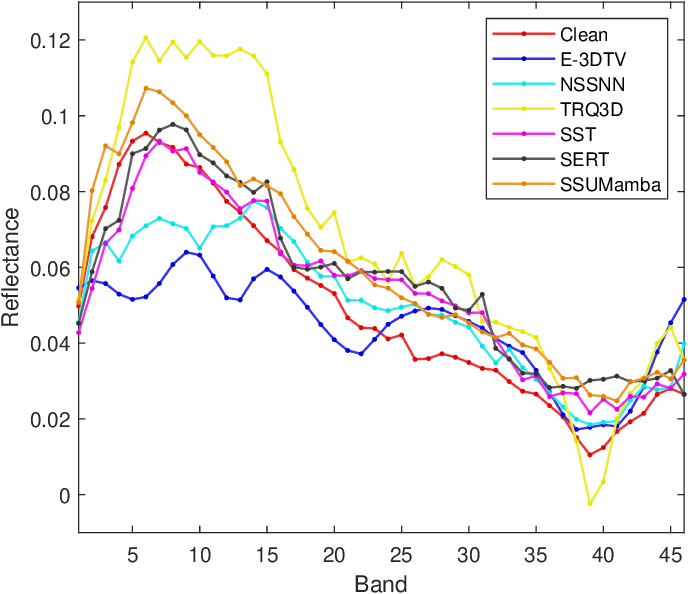}
	\caption{Reflectance of pixel (374,182) in the Houston 2018 HSI.}\label{fig:houston_pixel}
  \end{figure}

\subsubsection{Evaluation Indexs}

To quantitatively assess the denoising performance of all methods, we employed three reference metrics on synthetic datasets: peak signal-to-noise ratio (PSNR), structural similarity index (SSIM), and spectral angle mapper (SAM). Two no-reference metric, TOPIQ no reference (TOPIQ$\_$NR)~\cite{wang2023exploring} and Contrastive Language-Image Pre-training Image Quality Assessment (CLIPIQA+)~\cite{chen2024}, is employed on real-world HSIs. Higher PSNR, SSIM, TOPIQ$\_$NR, and CLIPIQA+ values indicate better denoising effectiveness, while smaller SAM values indicate superior performance.

\subsubsection{Comparison Methods}

We compared twelve methods in our study, comprising six model-based methods and six deep learning-based methods. The model-based methods include: BM4D~\cite{maggioni2012nonlocal}, MTSNMF~\cite{ye2014multitask}, NGMeet~\cite{He2020}, FastHyDe~\cite{zhuang2018fast}, LRTF$L_0$\cite{xiong2019}, and E-3DTV\cite{peng2020}. The deep learning-based methods consist of: T3SC~\cite{Bodrito2021}, MAC-Net~\cite{xiong2021mac}, NSSNN~\cite{guanyiman2022}, TRQ3D~\cite{Pang2022}, SST~\cite{li2022spatial}, and SERT~\cite{Li_2023_CVPR}. To ensure a fair comparison, we retrained one model for all the deep learning-based methods in each noise case.

\subsubsection{Architecture Design Details}
In  our SSUMamba, we incorporate 6 SSCS Mamba blocks with channels set as [32, 64, 64, 128, 128, 256]. Downsampling is enabled in blocks [2, 4, 6]. We employ the Adam optimizer with a learning rate of $3\times 10^{-4}$ to train the SSUMamba model for 45 epochs, with a reduction factor of 0.5 applied at the 20th and 35th epochs. The batch size is set to 14. The model is implemented using PyTorch and trained on a single NVIDIA GeForce RTX 4090 GPU.

\subsection{Comparison on Synthetic Datasets}

\begin{table*}[ht!]
	\caption{Comparison of Different Methods on Pavia City Center HSI. The Top Three Values Are Marked as \1{Red}, \2{Blue}, and \3{Green}.}\label{tab:pavia_center}
	\centering
	\resizebox{\linewidth}{!}{\tablesize{
		\begin{tabular}{c|c|c|c|c|c|c|c|c|c|c|c|c|c|c}
			\hline
		 &&\multicolumn{6}{c|}{\textbf{Model-based methods}}&\multicolumn{7}{c}{\textbf{Deep learning-based methods}}\\
		 \hline
		 \multirow{2}*{\makebox[0.02\textwidth][c]{Index}}&\multirow{2}*{\makebox[0.035\textwidth][c]{Noisy}}&\multirow{1}*{\makebox[0.035\textwidth][c]{BM4D}}&\multirow{1}*{\makebox[0.035\textwidth][c]{MTSNMF}}&\multirow{1}*{\makebox[0.035\textwidth][c]{NGMeet}}&\multirow{1}*{\makebox[0.035\textwidth][c]{FastHyDe}}&\multirow{1}*{\makebox[0.035\textwidth][c]{LRTF$L_0$}}&\multirow{1}*{\makebox[0.035\textwidth][c]{E-3DTV}}&\multirow{1}*{\makebox[0.035\textwidth][c]{T3SC}}&\multirow{1}*{\makebox[0.035\textwidth][c]{MAC-Net}}&\multirow{1}*{\makebox[0.035\textwidth][c]{NSSNN}}&\multirow{1}*{\makebox[0.035\textwidth][c]{TRQ3D}}&\multirow{1}*{\makebox[0.035\textwidth][c]{SST}}&\multirow{1}*{\makebox[0.035\textwidth][c]{SERT}}&\makebox[0.045\textwidth][c]{\textbf{SSUMamba}}\\
		 &&\cite{maggioni2012nonlocal}&\cite{ye2014multitask}&\cite{He2020}& \cite{zhuang2018fast}&\cite{xiong2019}& \cite{peng2020}& \cite{Bodrito2021}& \cite{xiong2021mac}& \cite{guanyiman2022}& \cite{Pang2022}& \cite{li2022spatial} & \cite{Li_2023_CVPR} & (ours)\\
		 \hline
   \makebox[0.02\textwidth][c]{PSNR$\uparrow$} & 13.46  & 21.70 &25.15  &23.68  &26.78 &26.49 &30.44 &28.69 &27.74 &\2{34.85}&28.23 &31.87 &\3{32.16}&\1{35.70}   \\
   \makebox[0.02\textwidth][c]{SSIM$\uparrow$} & .2181 &.5128 &.7397 &.7176  &.8361 &.8147 &.8941 &.8656 &.8724 &\1{.9581} &.8665 &.9234 &\3{.9544}&\2{.9546}   \\
   \makebox[0.02\textwidth][c]{SAM$\downarrow$} & .8893 &.5297 &.4094 &.4760  &.4040 &.3703 &\3{.1134} &.2135 &.3222 &\1{.1015}&.1961 &.1676 &.1384&\2{.1057}   \\
  \hline
  \end{tabular}}}
  \end{table*}

  \begin{figure*}[htbp]
	\centering
	\subfigure[{Clean}]{\label{fig:pavia_80_clean}\includegraphics[width=0.124\linewidth]{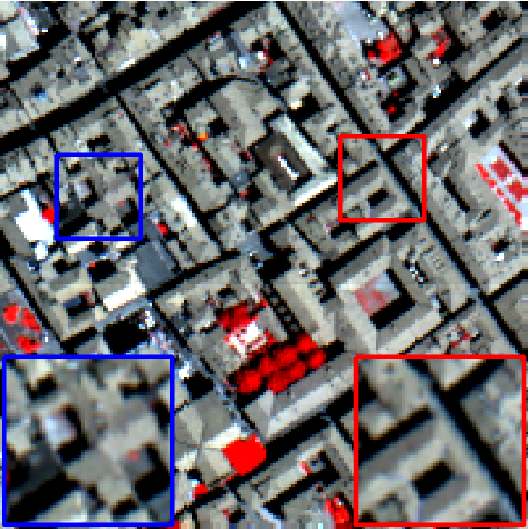}}
	\hspace{-2.1mm}
	\subfigure[{Noisy}]{\label{fig:pavia_80_noisy}\includegraphics[width=0.124\linewidth]{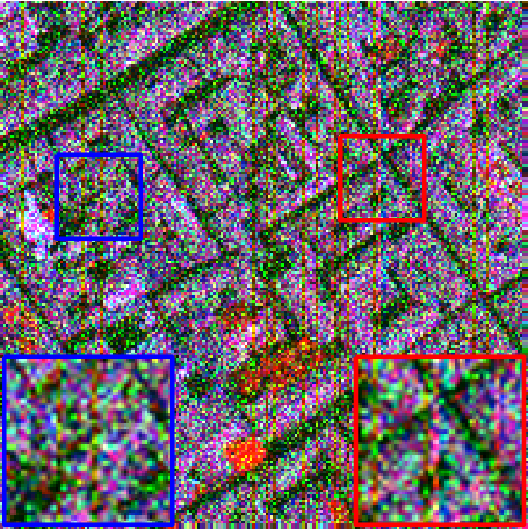}}
	\hspace{-2.1mm}
	\subfigure[{BM4D}~\cite{maggioni2012nonlocal}]{\label{fig:pavia_80_BM4D}\includegraphics[width=0.124\linewidth]{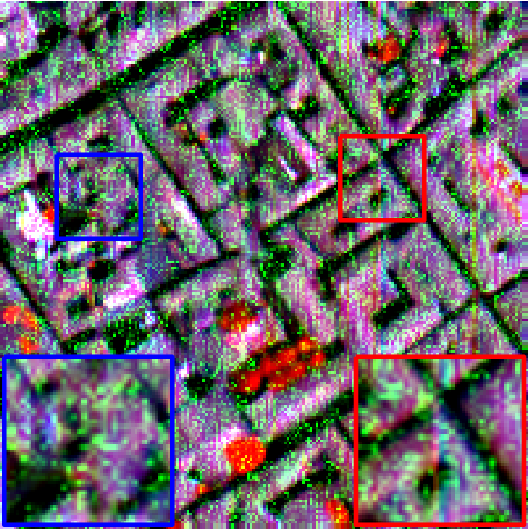}}
	\hspace{-2.1mm}
	\subfigure[{MTSNMF}~\cite{ye2014multitask}]{\label{fig:pavia_80_MTSNMF}\includegraphics[width=0.124\linewidth]{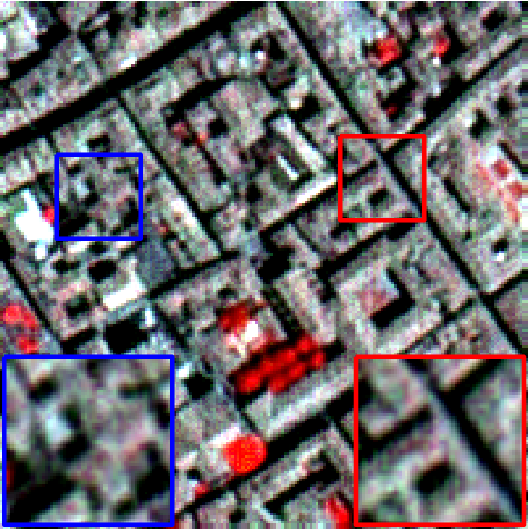}}
	\hspace{-2.1mm}
	\subfigure[{NGMeet}~\cite{He2020}]{\label{fig:pavia_80_NGMeet}\includegraphics[width=0.124\linewidth]{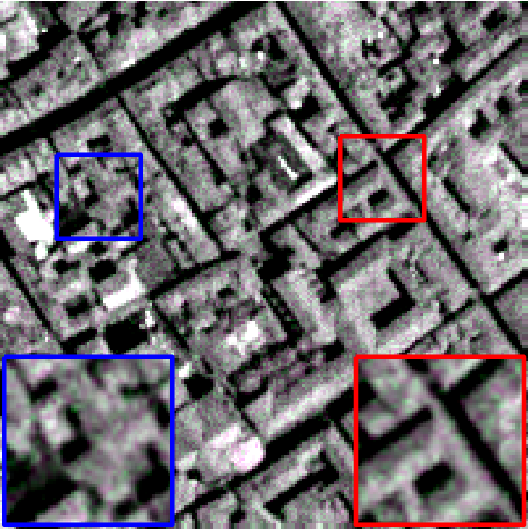}}
	\hspace{-2.1mm}
	\subfigure[{FastHyDe}~\cite{zhuang2018fast}]{\label{fig:pavia_80_FastHyDe}\includegraphics[width=0.124\linewidth]{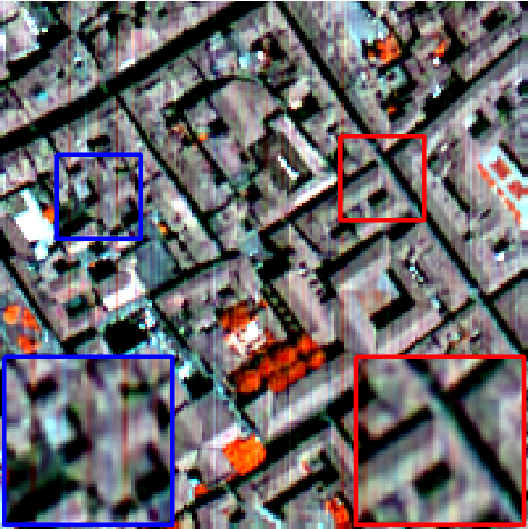}}
	\hspace{-2.1mm}
	\subfigure[{LRTF$L_0$}~\cite{xiong2019}]{\label{fig:pavia_80_lrtfl0}\includegraphics[width=0.124\linewidth]{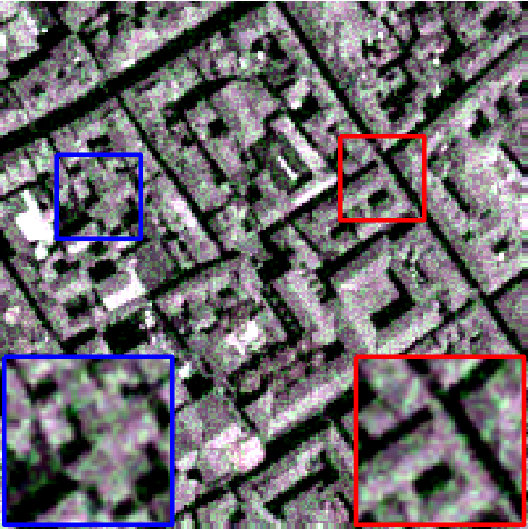}}
	\hspace{-2.1mm}
	\subfigure[{E-3DTV}~\cite{peng2020}]{\label{fig:pavia_80_e3dtv}\includegraphics[width=0.124\linewidth]{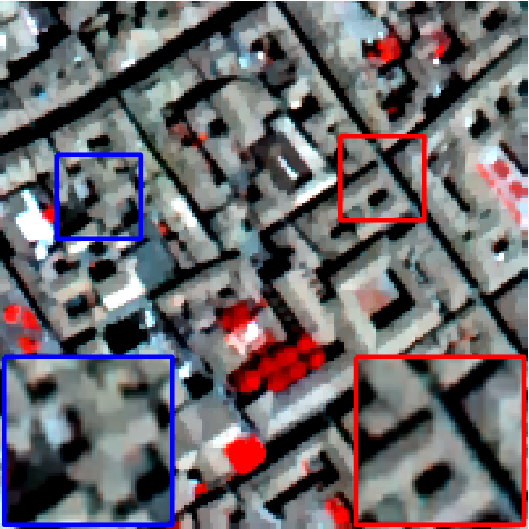}}\\
	\subfigure[{T3SC}~\cite{Bodrito2021}]{\label{fig:pavia_80_T3SC}\includegraphics[width=0.142\linewidth]{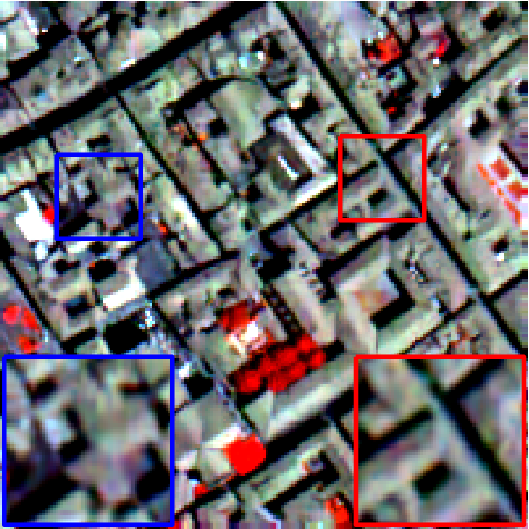}}
	\hspace{-2.1mm}
	\subfigure[{MAC-Net}~\cite{xiong2021mac}]{\label{fig:pavia_80_MAC-Net}\includegraphics[width=0.142\linewidth]{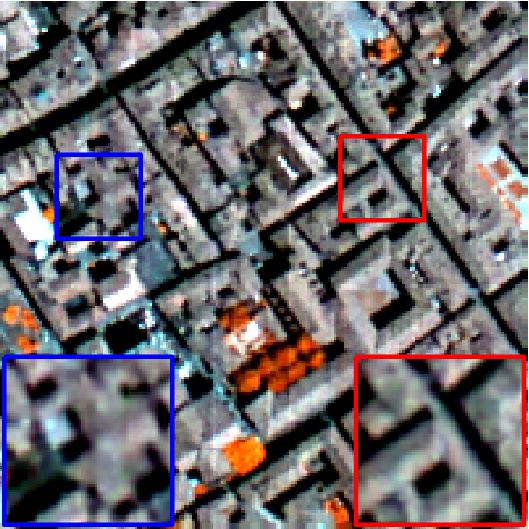}}
	\hspace{-2.1mm}
	\subfigure[NSSNN~\cite{guanyiman2022}]{\label{fig:pavia_80_NSSNN}\includegraphics[width=0.142\linewidth]{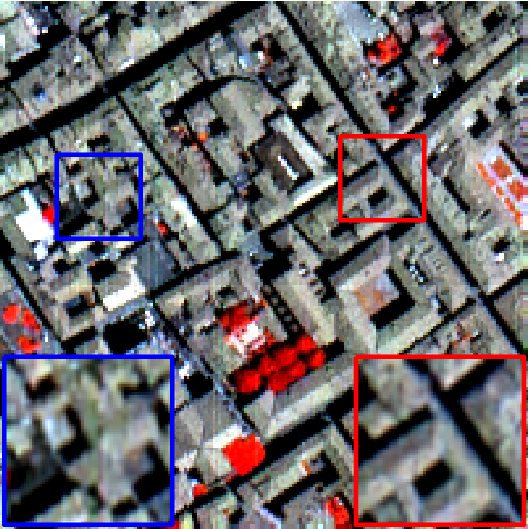}}
	\hspace{-2.1mm}
	\subfigure[{TRQ3D}~\cite{Pang2022}]{\label{fig:pavia_80_TRQ3D}\includegraphics[width=0.142\linewidth]{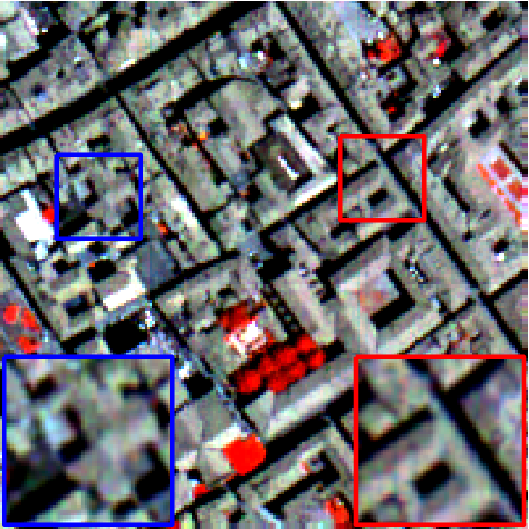}}
	\hspace{-2.1mm}
	\subfigure[{SST}~\cite{li2022spatial}]{\label{fig:pavia_80_SST}\includegraphics[width=0.142\linewidth]{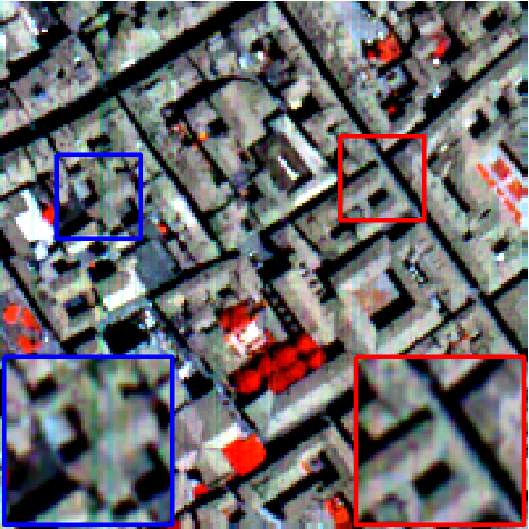}}
	\hspace{-2.1mm}
	\subfigure[SERT~\cite{Li_2023_CVPR}]{\label{fig:pavia_80_SERT}\includegraphics[width=0.142\linewidth]{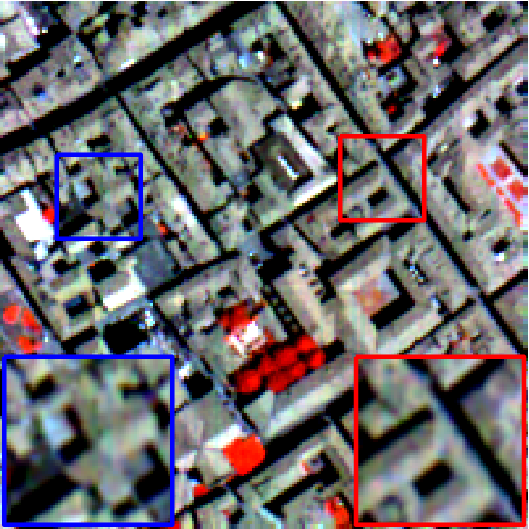}}
	\hspace{-2.1mm}
	\subfigure[{\textbf{SSUMamba}}]{\label{fig:pavia_80_SSRT}\includegraphics[width=0.142\linewidth]{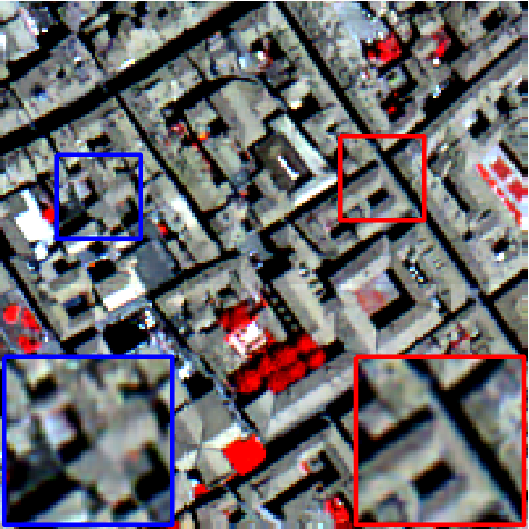}}
	   \caption{Denoising results on the Pavia City Center HSI with the mixture noise. The false-color images are generated by combining bands 65, 45, and 25.} \label{fig:pavia_80_visual}
 \end{figure*}

\begin{figure}[ht]
	\centering
	\includegraphics[width=0.85\linewidth]{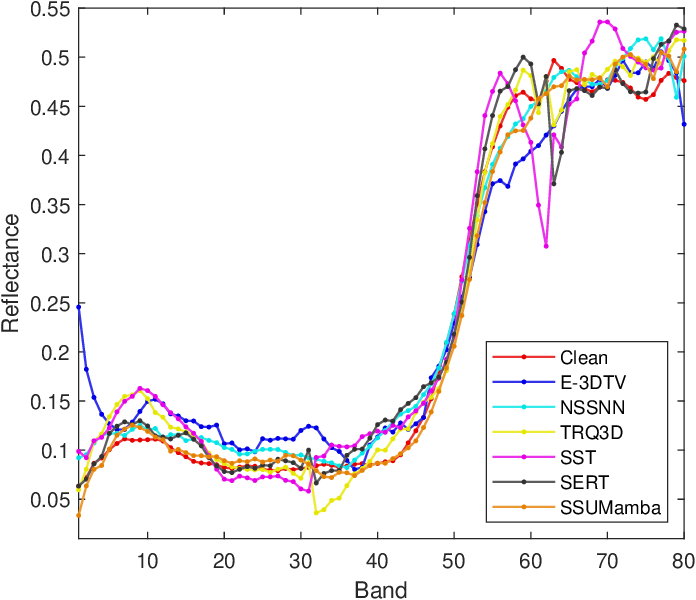}
	\caption{Reflectance of pixel (170,100) in the Pavia City Center HSI.}\label{fig:pavia_80_pixel}
  \end{figure}

\subsubsection{ICVL testing set}

Table~\ref{tab:icvl} presents the quantitative denoising results on the ICVL dataset, with the top three performers highlighted in bold \1{red}, bold \2{blue}, and bold \3{green}, respectively. FastHyDe and NGMeet demonstrate strong performance, particularly at lower noise levels, leveraging their ability to simultaneously model global spectral and nonlocal spatial correlations in HSIs. In cases of mixture noise, methods like LRTF$L_0$ and E-3DTV exhibit improved performance. However, model-based approaches tend to struggle in high noise and mixture noise conditions due to challenges in parameter optimization. T3SC and MAC-Net, integrating model-based techniques into data-driven learning, showcase robustness against Gaussian noise. NSSNN and TRQ3D utilize recurrent units to harness spectral correlations, demonstrating better adaptability to mixed noise due to their data-driven learning approach. SST outperforms NSSNN and TRQ3D by using spectral and spatial transformer-based modules to enhance the long-range correlations. SERT stands out among other methods except for our SSUMamba, leveraging spatial cross-rectangle self-attention and low-rank-based spectral enhancement modules. Notably, SSUMamba consistently outperforms other methods across all noise levels and metrics, showcasing superior denoising capabilities, structural preservation, and spectral fidelity. This superiority is attributed to its ability to effectively model spatial-spectral correlation and ensure spatial-spectral continuity for HSI sequence generation.

\subsubsection{Houston 2018 HSI}

The quantitative results for the Houston 2018 HSI are presented in Table~\ref{tab:houston}. While all deep learning-based methods excel in learning mixture noise cases from the ICVL dataset and outperform model-based methods, they face significant challenges when applied to the Houston 2018 HSI due to substantial differences between the datasets. As a result, their denoising performance declines noticeably. Despite these challenges, our proposed SSUMamba, leveraging an advanced scanning scheme that considers both spatial-spectral correlation and continuity during long-range modeling, substantially outperforms transformer-based approaches. 


Fig.~\ref{fig:houston_visual} illustrates the false-color images recovered from all comparison methods. Conventional model-based approaches, designed for Gaussian noise, struggle to eliminate artifacts such as strips. However, E-3DTV stands out by employing the $L_1$-norm for sparse noise modeling. TRQ3D retains Gaussian noise and stripe artifacts, while SST exhibits color distortion, indicating inaccuracies in spectral fidelity relative to the noisy image. NSSNN and SERT achieve better visualization, but there are still some deadlines and strips. In contrast, our SSUMamba effectively utilizes SSCS in Mamba blocks to capture long-range correlations and achieves the best visual performance.

\begin{table*}[htbp]
	\caption{Comparison of Different Methods on Gaofen-5 Wuhan HSI. The Top Three Values Are Marked as \1{Red}, \2{Blue}, and \3{Green}.}\label{tab:wuhan}
	\centering
	\resizebox{\linewidth}{!}{\tablesize{
		\begin{tabular}{c|c|c|c|c|c|c|c|c|c|c|c|c|c|c}
			\hline
		 &&\multicolumn{6}{c|}{\textbf{Model-based methods}}&\multicolumn{7}{c}{\textbf{Deep learning-based methods}}\\
		 \hline
		 \multirow{2}*{\makebox[0.02\textwidth][c]{Index}}&\multirow{2}*{\makebox[0.035\textwidth][c]{Noisy}}&\multirow{1}*{\makebox[0.035\textwidth][c]{BM4D}}&\multirow{1}*{\makebox[0.035\textwidth][c]{MTSNMF}}&\multirow{1}*{\makebox[0.035\textwidth][c]{NGMeet}}&\multirow{1}*{\makebox[0.035\textwidth][c]{FastHyDe}}&\multirow{1}*{\makebox[0.035\textwidth][c]{LRTF$L_0$}}&\multirow{1}*{\makebox[0.035\textwidth][c]{E-3DTV}}&\multirow{1}*{\makebox[0.035\textwidth][c]{T3SC}}&\multirow{1}*{\makebox[0.035\textwidth][c]{MAC-Net}}&\multirow{1}*{\makebox[0.035\textwidth][c]{NSSNN}}&\multirow{1}*{\makebox[0.035\textwidth][c]{TRQ3D}}&\multirow{1}*{\makebox[0.035\textwidth][c]{SST}}&\multirow{1}*{\makebox[0.035\textwidth][c]{SERT}}&\makebox[0.045\textwidth][c]{\textbf{SSUMamba}}\\
		 &&\cite{maggioni2012nonlocal}&\cite{ye2014multitask}&\cite{He2020}& \cite{zhuang2018fast}&\cite{xiong2019}& \cite{peng2020}& \cite{Bodrito2021}& \cite{xiong2021mac}& \cite{guanyiman2022}& \cite{Pang2022}& \cite{li2022spatial} & \cite{Li_2023_CVPR} & (ours)\\
		 \hline
TOPIQ$\_$NR$\uparrow$ &.3826&.3828&.3745&.3650&\3{.3887}&.3877&.3815&\2{.4274}&.3477&.3403&.3840&.3783&.3824&\1{.4969}\\
CLIPIQA+$\uparrow$    &.5824&.5827&.6005&.4102&.6001&.6021&\3{.6055}&.5735&.5549&.5901&.5956&.5874&\2{.6081}&\1{.6481}\\
	\hline	
	\end{tabular}}}
\end{table*}

\begin{figure*}[ht!]
	\centering
	\subfigure[Noisy]{\label{fig:wuhan_noisy}\includegraphics[width=0.142\linewidth]{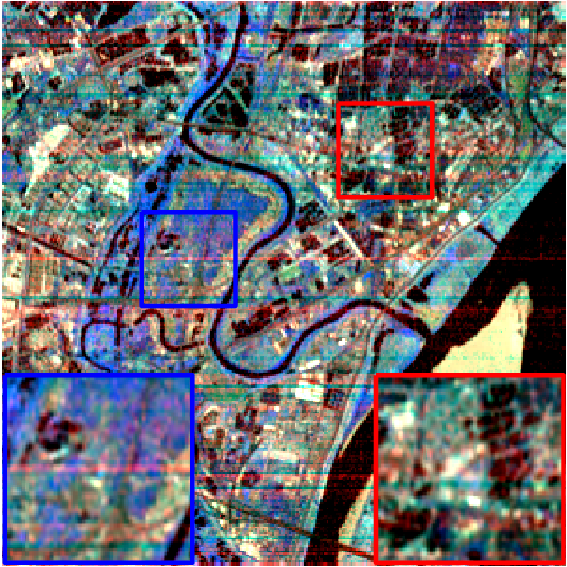}}
	\hspace{-2.1mm}
	\subfigure[BM4D \cite{maggioni2012nonlocal}]{\label{fig:wuhan_BM4D}\includegraphics[width=0.142\linewidth]{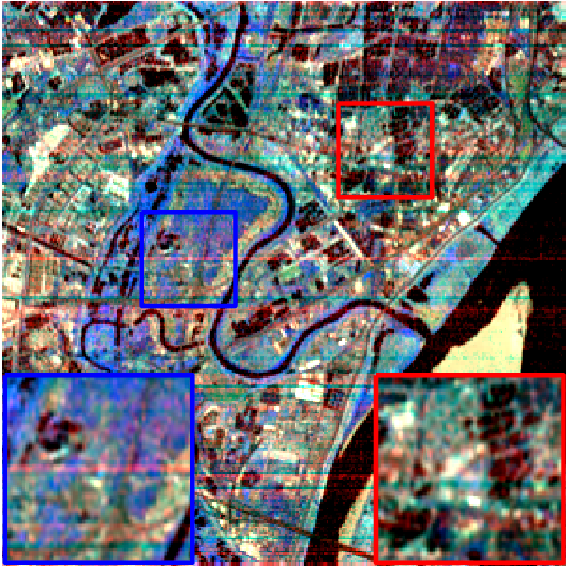}}
	\hspace{-2.1mm}
	\subfigure[MTSNMF \cite{ye2014multitask}]{\label{fig:wuhan_MTSNMF}\includegraphics[width=0.142\linewidth]{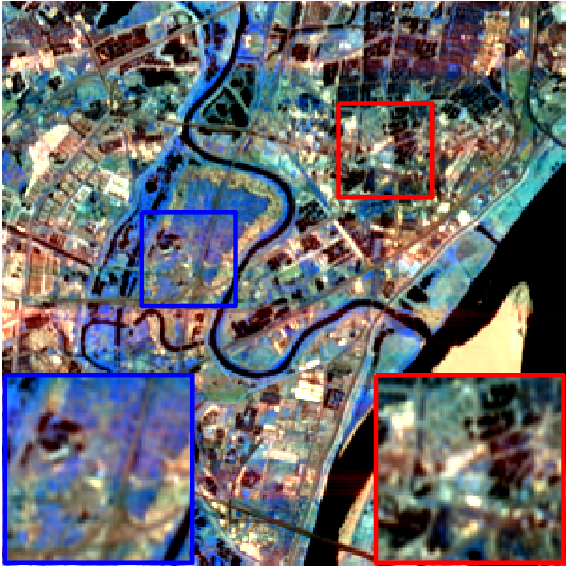}}
	\hspace{-2.1mm}
	\subfigure[NGMeet \cite{He2020}]{\label{fig:wuhan_NGMeet}\includegraphics[width=0.142\linewidth]{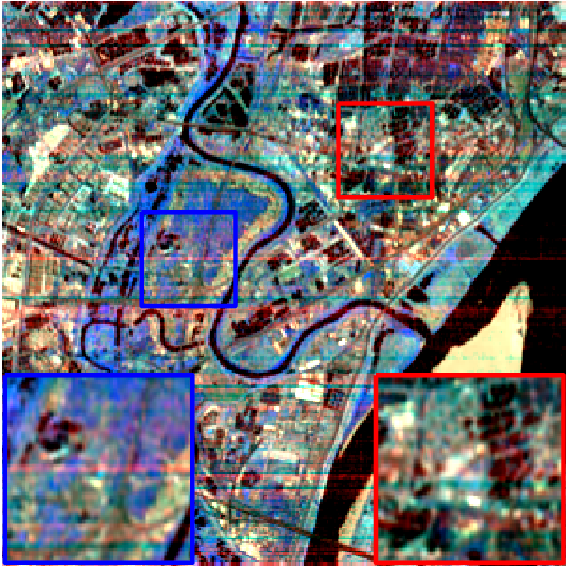}}
	\hspace{-2.1mm}
	\subfigure[FastHyDe \cite{zhuang2018fast}]{\label{fig:wuhan_FastHyDe}\includegraphics[width=0.142\linewidth]{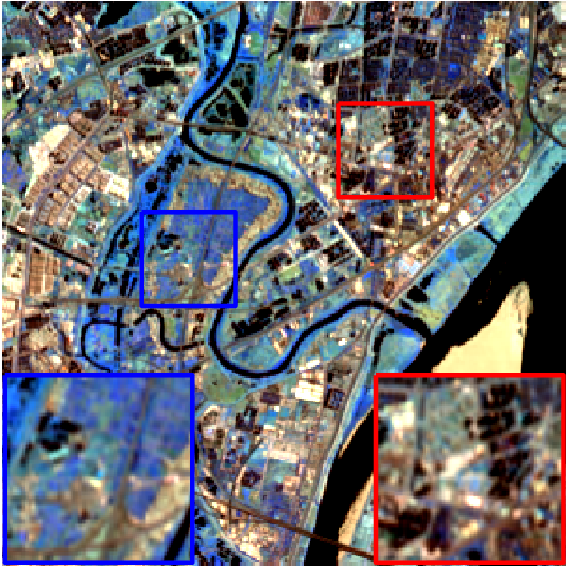}}
	\hspace{-2.1mm}
	\subfigure[LRTF$L_0$ \cite{xiong2019}]{\label{fig:wuhan_lrtfl0}\includegraphics[width=0.142\linewidth]{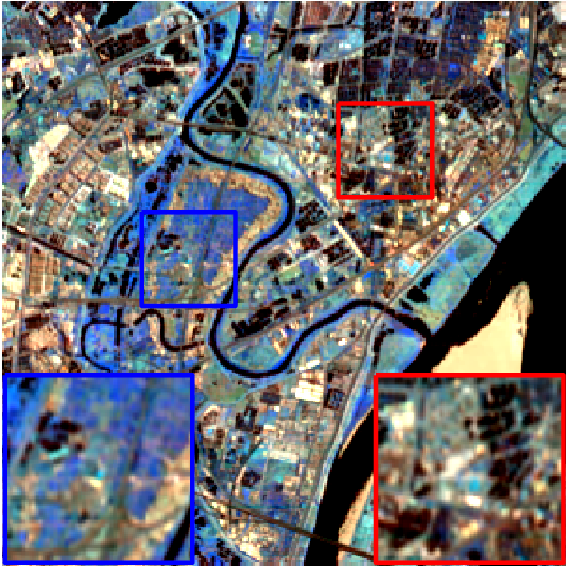}}
	\hspace{-2.1mm}
	\subfigure[E-3DTV \cite{peng2020}]{\label{fig:wuhan_e3dtv}\includegraphics[width=0.142\linewidth]{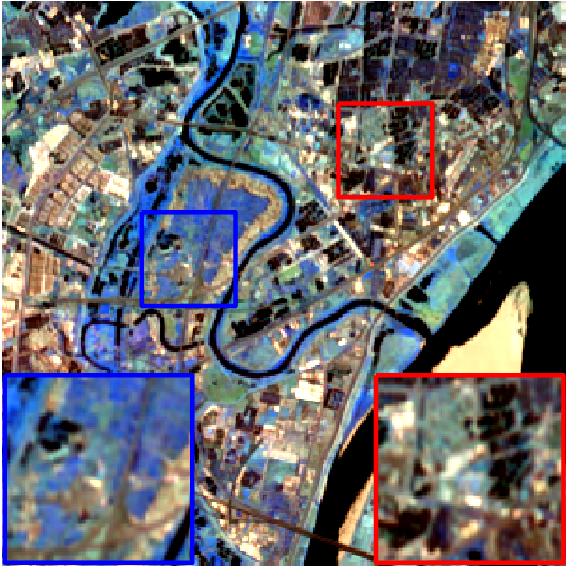}}\\
	\hspace{-2.1mm}
	\subfigure[T3SC \cite{Bodrito2021}]{\label{fig:wuhan_T3SC}\includegraphics[width=0.142\linewidth]{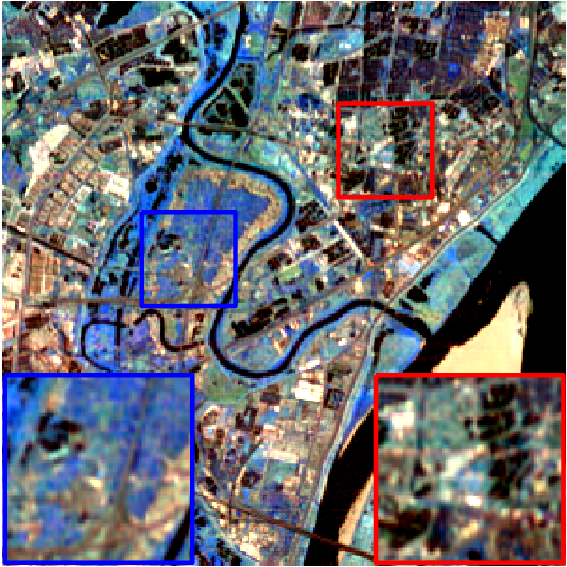}}
	\hspace{-2.1mm}
	\subfigure[MAC-Net \cite{xiong2021mac}]{\label{fig:wuhan_MAC-Net}\includegraphics[width=0.1420\linewidth]{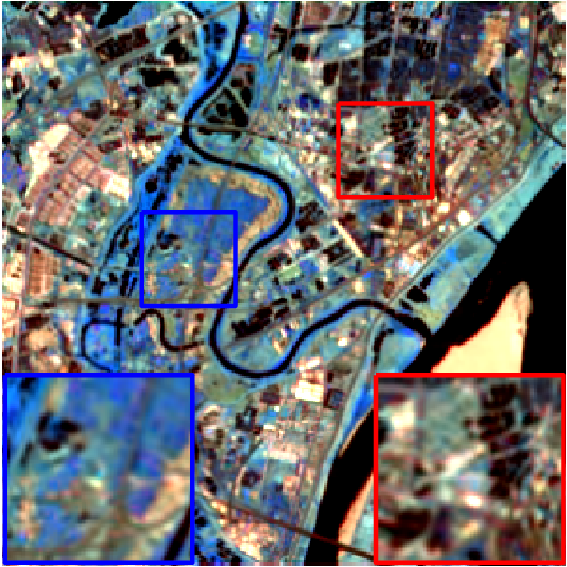}}
	\hspace{-2.1mm}
	\subfigure[NSSNN \cite{guanyiman2022}]{\label{fig:wuhan_NSSNN}\includegraphics[width=0.1420\linewidth]{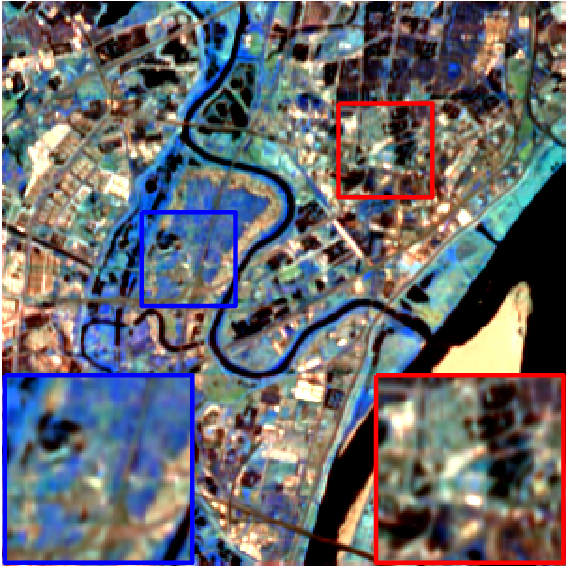}}
	\hspace{-2.1mm}
	\subfigure[TRQ3D \cite{Pang2022}]{\label{fig:wuhan_TRQ3D}\includegraphics[width=0.1420\linewidth]{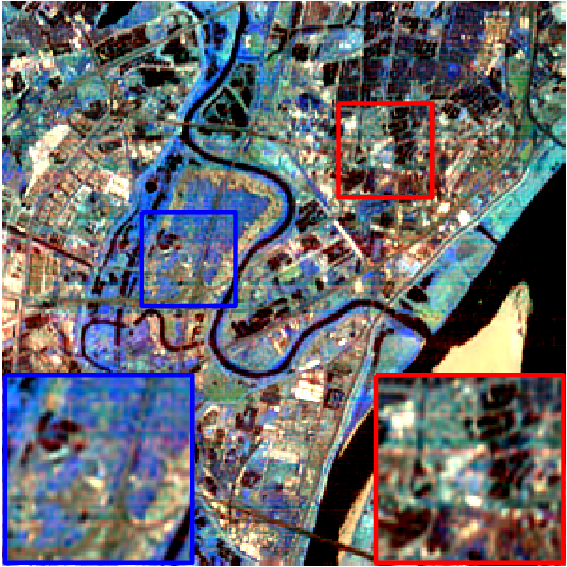}}
	\hspace{-2.1mm}
	\subfigure[SST \cite{li2022spatial}]{\label{fig:wuhan_SST}\includegraphics[width=0.1420\linewidth]{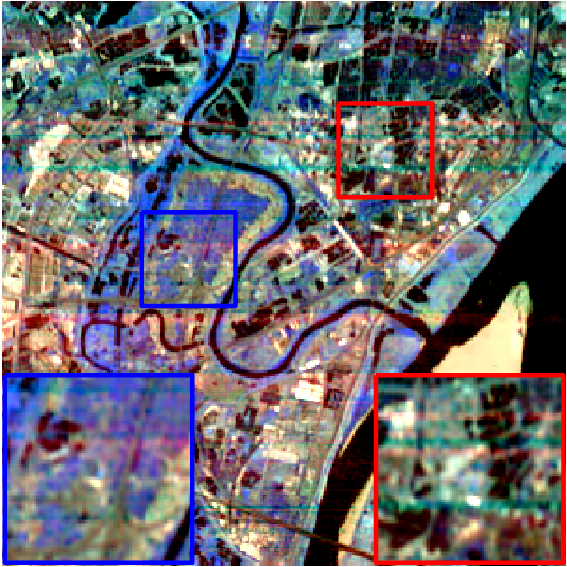}}
	\hspace{-2.1mm}
	\subfigure[SERT \cite{Li_2023_CVPR}]{\label{fig:wuhan_SERT}\includegraphics[width=0.1420\linewidth]{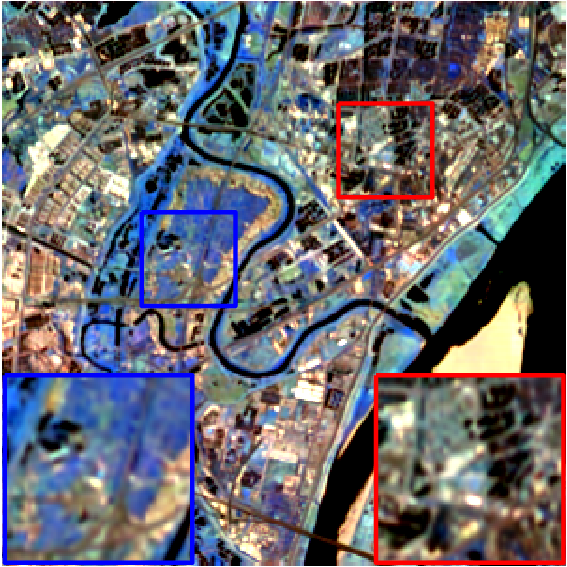}}
	\hspace{-2.1mm}
	\subfigure[\textbf{SSUMamba}]{\label{fig:wuhan_SSRT}\includegraphics[width=0.1420\linewidth]{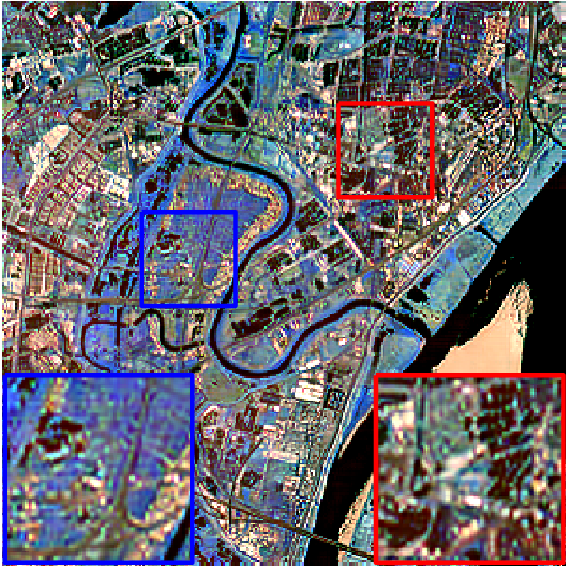}}
  \caption{Denoising results on the Gaofen-5 Wuhan HSI. The false-color images are generated by combining bands 152, 83, and 33.} \label{fig:wuhan_visual}
 \end{figure*}

\begin{figure}[ht]
	\centering
	\includegraphics[width=0.85\linewidth]{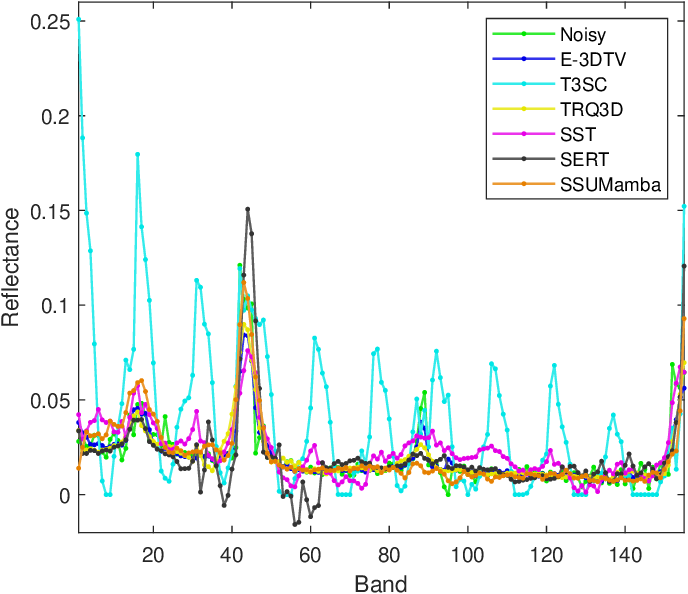}
	   \caption{Reflectance of pixel (248,273) in the Gaofen-5 Wuhan HSI.} \label{fig:wuhan_pixel}
 \end{figure}

The spectral reflectances are shown in Fig.~\ref{fig:houston_pixel}, where the mixture noise causes significant spectral distortion. To enhance intuition, we only present the spectral reflectances of clean images, noisy images, and the top six  methods: E-3DTV, NSSNN, TRQ3D, SST, SERT, and our SSUMamba.
The model-based E-3DTV produces smooth spectral reflectance but deviates from the clean HSI. TRQ3D, and SERT encounter difficulties due to the differing number of bands in the Houston 2018 HSI compared to the ICVL training set, leading to incomplete modeling of global spectral correlation and less accurate matching of spectral reflectances with the clean HSI. SST, which employs a transformer-based module to model spectral information, achieves spectral reflectance closer to the clean HSI than the other comparison methods.
Our proposed SSUMamba excels by capturing long-range spectral correlations and continuity, effectively recovering spectral reflectance and demonstrating superior performance in spectral modeling compared to all other methods.

\subsubsection{Pavia City Center HSI}
Table~\ref{tab:pavia_center} offers a detailed comparison of various denoising methods applied to the Pavia City Center HSI. Model-based E-3DTV leverages sparsity calculations on gradient map subspaces across all HSI bands and demonstrates commendable performance in noise reduction, which is particularly evident in the SAM metric. NSSNN achieves better performance in the comparison methods by using a gated recurrent network to capture spectral correlation and a criss-cross attention to model nonlocal spatial correlation. Our proposed SSUMamba utilizes advanced scanning schemes that consider spatial-spectral correlation and continuity during long-range modeling. This results in superior denoising capabilities, as evidenced by its best PSNR and close SSIM and SAM metrics with those of NSSNN.

Fig.~\ref{fig:pavia_80_visual} presents false-color images recovered by all comparison methods. While model-based methods, tailored for Gaussian noise, struggle to eliminate artifacts like strips and distortions in mixture noise, E-3DTV stands out as the top performer among them. Despite its effectiveness, E-3DTV still exhibits blurring due to limited spatial correlation modeling capabilities. NSSNN did not completely remove strip noises. Transformer-based approaches, such as TRQ3D, SST, and SERT, encounter challenges in reconstruction due to spectral variations between the Pavia City Center HSI and the ICVL training set. In contrast, our SSUMamba, adept at capturing long-range spatial-spectral correlations, achieves highly accurate reconstructions with minimal artifacts. Fig.~\ref{fig:pavia_80_pixel} displays the spectral reflectances. SSUMamba effectively leverages spectral continuity in long-range SSM modeling to recover spectral reflectance, resulting in the closest match to the clean HSI among all methods.

\begin{table*}[htbp]
	\caption{Comparison of Different Methods on Earth Observing-1 HSI. The Top Three Values Are Marked as \1{Red}, \2{Blue}, and \3{Green}.}\label{tab:eo1}
	\centering
	\resizebox{\linewidth}{!}{\tablesize{
		\begin{tabular}{c|c|c|c|c|c|c|c|c|c|c|c|c|c|c}
			\hline
		 &&\multicolumn{6}{c|}{\textbf{Model-based methods}}&\multicolumn{7}{c}{\textbf{Deep learning-based methods}}\\
		 \hline
		 \multirow{2}*{\makebox[0.02\textwidth][c]{Index}}&\multirow{2}*{\makebox[0.035\textwidth][c]{Noisy}}&\multirow{1}*{\makebox[0.035\textwidth][c]{BM4D}}&\multirow{1}*{\makebox[0.035\textwidth][c]{MTSNMF}}&\multirow{1}*{\makebox[0.035\textwidth][c]{NGMeet}}&\multirow{1}*{\makebox[0.035\textwidth][c]{FastHyDe}}&\multirow{1}*{\makebox[0.035\textwidth][c]{LRTF$L_0$}}&\multirow{1}*{\makebox[0.035\textwidth][c]{E-3DTV}}&\multirow{1}*{\makebox[0.035\textwidth][c]{T3SC}}&\multirow{1}*{\makebox[0.035\textwidth][c]{MAC-Net}}&\multirow{1}*{\makebox[0.035\textwidth][c]{NSSNN}}&\multirow{1}*{\makebox[0.035\textwidth][c]{TRQ3D}}&\multirow{1}*{\makebox[0.035\textwidth][c]{SST}}&\multirow{1}*{\makebox[0.035\textwidth][c]{SERT}}&\makebox[0.045\textwidth][c]{\textbf{SSUMamba}}\\
		 &&\cite{maggioni2012nonlocal}&\cite{ye2014multitask}&\cite{He2020}& \cite{zhuang2018fast}&\cite{xiong2019}& \cite{peng2020}& \cite{Bodrito2021}& \cite{xiong2021mac}& \cite{guanyiman2022}& \cite{Pang2022}& \cite{li2022spatial} & \cite{Li_2023_CVPR} & (ours)\\
		 \hline
TOPIQ$\_$NR$\uparrow$ &.5193&.5184&.4263&.4191&\3{.5235}&.5177&.4838&\2{.5258}&.5231&.4175&.3536&.4908&.4654&\1{.6074}\\
CLIPIQA+$\uparrow$    &.4935&.5011&.4838&.3951&.5187&\3{.5204}&\2{.5355}&.4892&.4964&.4898&.4473&.5169&.5161&\1{.5559}\\
	\hline	
	\end{tabular}}}
\end{table*}

\begin{figure*}[htbp]
	\centering
	\subfigure[{Noisy}]{\label{fig:eo1_noisy}\includegraphics[width=0.142\linewidth]{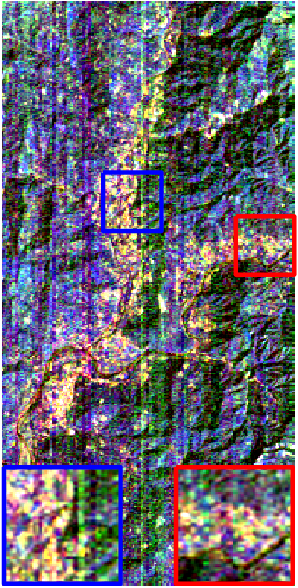}}
	\hspace{-2.1mm}
	\subfigure[{BM4D} \cite{maggioni2012nonlocal}]{\label{fig:eo1_BM4D}\includegraphics[width=0.142\linewidth]{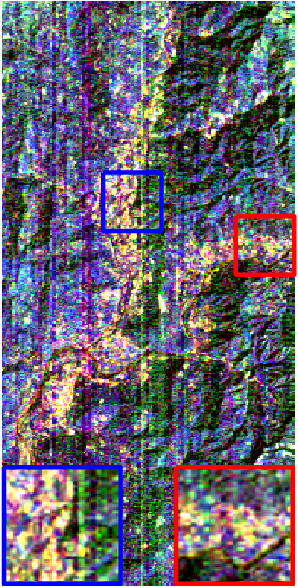}}
	\hspace{-2.1mm}
	\subfigure[{MTSNMF} \cite{ye2014multitask}]{\label{fig:eo1_MTSNMF}\includegraphics[width=0.142\linewidth]{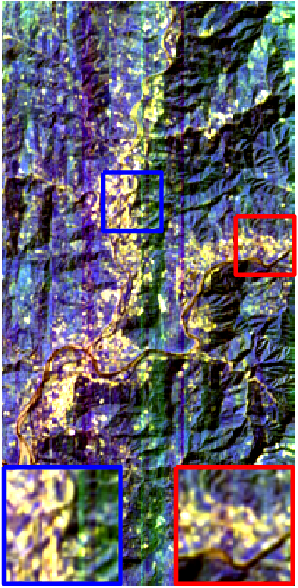}}
	\hspace{-2.1mm}
	\subfigure[{NGMeet} \cite{He2020}]{\label{fig:eo1_NGMeet}\includegraphics[width=0.142\linewidth]{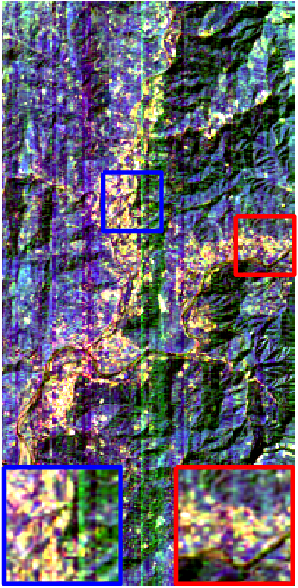}}
	\hspace{-2.1mm} 
	\subfigure[{FastHyDe} \cite{zhuang2018fast}]{\label{fig:eo1_FastHyDe}\includegraphics[width=0.142\linewidth]{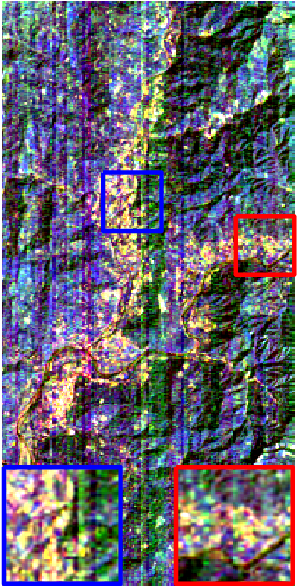}}
	\hspace{-2.1mm}
	\subfigure[{LRTF$L_0$} \cite{xiong2019}]{\label{fig:eo1_lrtfl0}\includegraphics[width=0.142\linewidth]{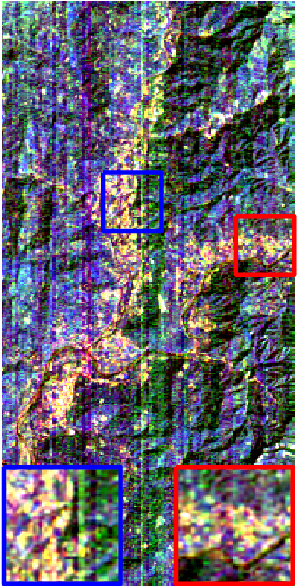}}
	\hspace{-2.1mm}
	\subfigure[{E-3DTV} \cite{peng2020}]{\label{fig:eo1_e3dtv}\includegraphics[width=0.142\linewidth]{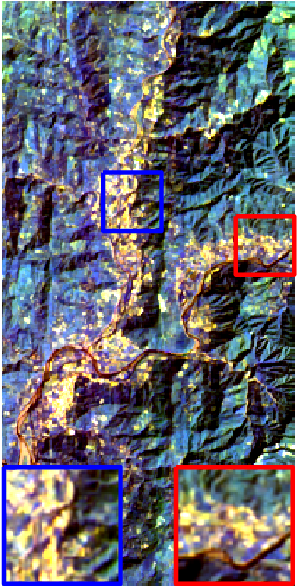}}
	\hspace{-2.1mm}
	\subfigure[{T3SC} \cite{Bodrito2021}]{\label{fig:eo1_T3SC}\includegraphics[width=0.1420\linewidth]{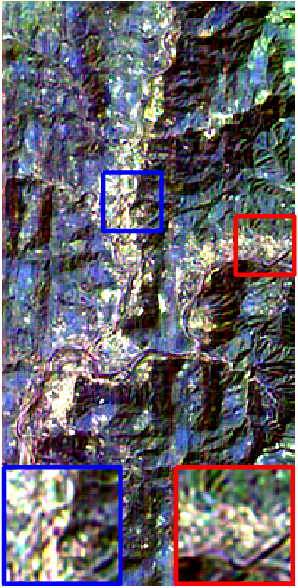}}
	\hspace{-2.1mm}
	\subfigure[{MAC-Net} \cite{xiong2021mac}]{\label{fig:eo1_MAC-Net}\includegraphics[width=0.1421\linewidth]{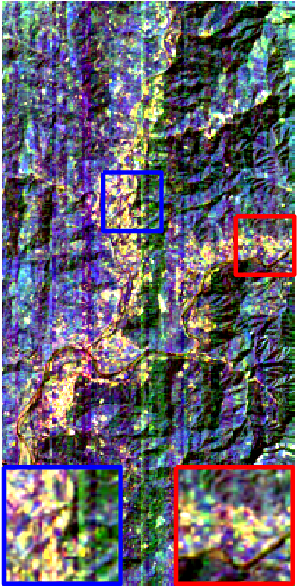}}
	\hspace{-2.1mm}
	\subfigure[NSSNN \cite{li2022spatial}]{\label{fig:eo1_NSSNN}\includegraphics[width=0.1420\linewidth]{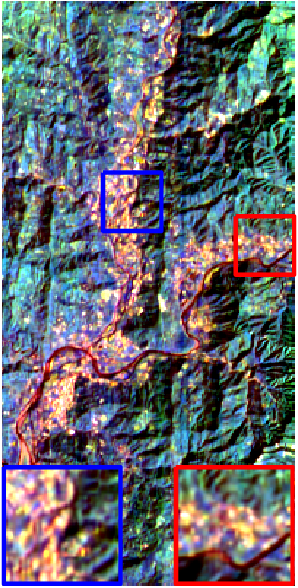}}
	\hspace{-2.1mm}
	\subfigure[{TRQ3D} \cite{Pang2022}]{\label{fig:eo1_TRQ3D}\includegraphics[width=0.1420\linewidth]{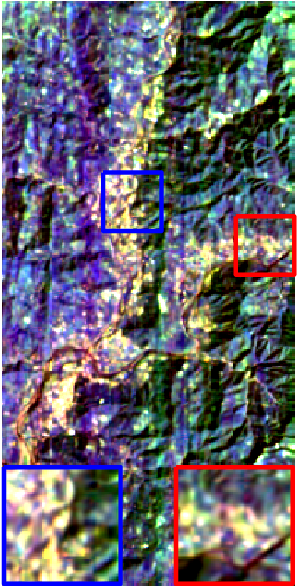}}
	\hspace{-2.1mm}
	\subfigure[{SST} \cite{li2022spatial}]{\label{fig:eo1_SST}\includegraphics[width=0.1420\linewidth]{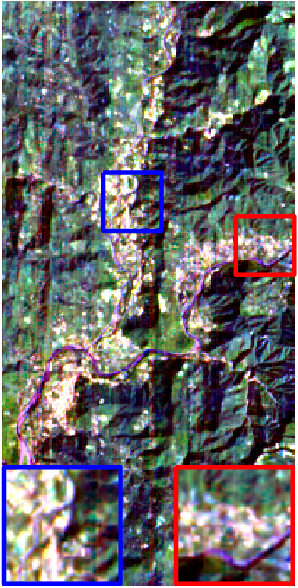}}
	\hspace{-2.1mm}
	\subfigure[SERT \cite{li2022spatial}]{\label{fig:eo1_SERT}\includegraphics[width=0.1420\linewidth]{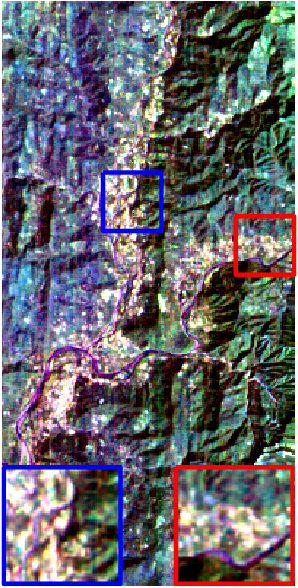}}
	\hspace{-2.1mm}
	\subfigure[{\textbf{SSUMamba}}]{\label{fig:eo1_SSRT}\includegraphics[width=0.1420\linewidth]{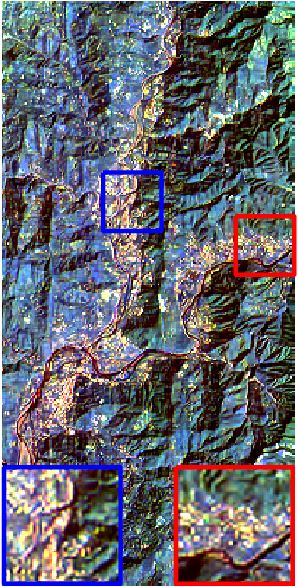}}
  \caption{{Denoising results on the Earth  Observing-1 HSI. The false-color images are generated by combining bands 163, 96, and 30.}} \label{fig:eo1_visual}
  \end{figure*}

\begin{figure}[!htbp]
	\centering
	\includegraphics[width=0.85\linewidth]{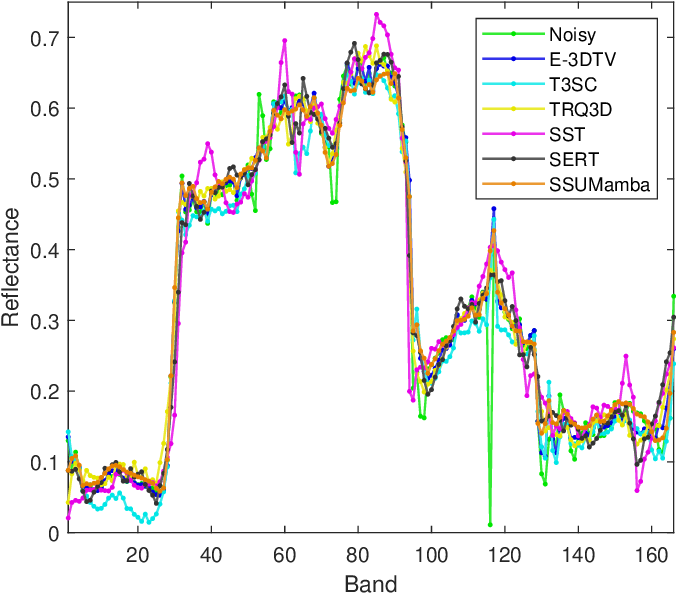}
	   \caption{Reflectance of pixel (151,124) in the Earth  Observing-1 HSI.} \label{fig:eo1_pixel}
    \vspace{-0.2cm}
  \end{figure}

\subsection{Comparison on Real-World HSIs}

We carried out extra experiments on real-world remote-sensing HSIs to thoroughly evaluate the effectiveness of denoising across various methods. Since there is no clean HSI in real-world comparisons, we provide no-reference measures, \emph{i.e.}, TOPIQ$\_$NR and CLIPIQA+, and qualitative comparative analysis to assess the denoising results.

\subsubsection{Gaofen-5 Wuhan HSI}

Table~\ref{tab:wuhan} presents the quantitative comparison of various methods on the Gaofen-5 Wuhan HSI. 
Our proposed SSUMamba outperforms all other methods, demonstrating its ability to restore both high-level and low-level information, \emph{i.e.}, construction and texture of the image, in real-world HSI denoising.
The visualization results on the Gaofen-5 Wuhan HSI are presented in Fig.~\ref{fig:wuhan_visual}, where the noisy image is affected by impulse noise and strips. Among the model-based methods, FastHyDe and LRTF$L_0$ demonstrate superior performance, attributed to their effectiveness in addressing mixture noise. T3SC and MAC-Net successfully mitigate most of the strips present in the Gaofen-5 Wuhan HSI. Notably, T3SC yields sharper results by integrating spectral and spatial information within a trainable sparse coding framework. However, transformer-based methods like TRQ3D and SST struggle to completely eliminate all the strips, indicating limitations in adaptability to the specific imaging environment. NSSNN and SERT can remove most of the strips but result in blurred HSIs with indistinct edges. In contrast, our SSUMamba approach leverages its capability to effectively capture spatial-spectral correlation and encode the 3-D structure of HSIs, resulting in highly accurate reconstructions with sharper edges compared to other methods. The recovered spectral reflectances are depicted in Fig.~\ref{fig:wuhan_pixel}. Considering that T3SC shows better performance in real-world HSI denoising, we therefore add T3SC instead of NSSNN in the comparison. By harnessing long-range spectral correlation learning coupled with the 3-D characteristics of HSIs, SSUMamba demonstrates a superior ability to recover spectral reflectance with greater accuracy. The smoother curves closely resembling the original shape indicate its effectiveness in mitigating the effects of impulse noise and strips on spectral fidelity.

\begin{table*}[!htbp]
	\centering
		\caption{Ablation Study on The Effect of Introduced Components.}
		\label{tab:ab_module}
	   \resizebox{1\linewidth}{!}{
		\begin{tabular}{ c |c |c|c |c |c|c |c |c|c |c |c}
		   \hline
	 \multirow{2}{*}{\makebox[0.08\textwidth][c]{SSCS}}&\multirow{2}{*}{\makebox[0.08\textwidth][c]{Bidirectional}}&\multirow{2}{*}{\makebox[0.08\textwidth][c]{Residual Block}}&\multicolumn{3}{c|}{ICVL}&\multicolumn{3}{c|}{Houston 2018 HSI}&\multicolumn{3}{c}{Pavia City Center HSI}\\
	 &&&\makebox[0.05\textwidth][c]{PSNR$\uparrow$}&\makebox[0.05\textwidth][c]{SSIM$\uparrow$}&\makebox[0.05\textwidth][c]{SAM$\downarrow$}&\makebox[0.05\textwidth][c]{PSNR$\uparrow$}&\makebox[0.05\textwidth][c]{SSIM$\uparrow$}&\makebox[0.05\textwidth][c]{SAM$\downarrow$}&\makebox[0.05\textwidth][c]{PSNR$\uparrow$}&\makebox[0.05\textwidth][c]{SSIM$\uparrow$}&\makebox[0.05\textwidth][c]{SAM$\downarrow$}\\
		\hline
\Checkmark&\Checkmark &\Checkmark&\textbf{43.07}&\textbf{.9726}&.0710&\textbf{34.74}&\textbf{.9452}&\textbf{.0993}&35.70&\textbf{.9546}&.1057\\
\XSolidBrush&\Checkmark&\Checkmark&42.73&.9710&.0689&34.40&.9444&.1056&\textbf{35.76}&.9507&\textbf{.1036}\\
\Checkmark&\XSolidBrush&\Checkmark&42.74&.9701&\textbf{.0672}&34.51&.9435&.1051&35.63&.9528&.1139\\
\Checkmark&\Checkmark&\XSolidBrush&42.69&.9715&.0700&34.44&.9436&.1054&35.29&.9533&.1089\\
		\hline
	 \end{tabular}}
	 \end{table*}
	 
	 \begin{table*}[!htbp]
		\centering
		\caption{Ablation Study on The Impact of Network Width (Channels).}
		\label{tab:ab_width}
		 \resizebox{1\linewidth}{!}{
		\begin{tabular}{ c |c|c |c| c|c| c| c|c| c| c|c| c| c}
		\hline
		\multirow{2}{*}{\makebox[0.05\textwidth][c]{Channels}}&\multirow{2}{*}{\makebox[0.05\textwidth][c]{\#Param.}}&\multirow{2}{*}{\makebox[0.05\textwidth][c]{Memory}}&\multirow{2}{*}{\makebox[0.05\textwidth][c]{FLOPS}}&\multirow{2}{*}{\makebox[0.05\textwidth][c]{Time}}&\multicolumn{3}{c|}{ICVL}&\multicolumn{3}{c|}{Houston 2018 HSI}&\multicolumn{3}{c}{Pavia City Center HSI}\\
		&&&&&\makebox[0.045\textwidth][c]{PSNR$\uparrow$}&\makebox[0.045\textwidth][c]{SSIM$\uparrow$}&\makebox[0.045\textwidth][c]{SAM$\downarrow$}&\makebox[0.045\textwidth][c]{PSNR$\uparrow$}&\makebox[0.045\textwidth][c]{SSIM$\uparrow$}&\makebox[0.045\textwidth][c]{SAM$\downarrow$}&\makebox[0.045\textwidth][c]{PSNR$\uparrow$}&\makebox[0.045\textwidth][c]{SSIM$\uparrow$}&\makebox[0.045\textwidth][c]{SAM$\downarrow$}\\
		\hline
		32&10.4M&1,802M		&96T&1.06s&43.07	&.9726	&.0710	&34.74	&.9452	&.0993	&35.70	&.9546	&.1057	\\
		24&5.9M&1,386M		&70T&0.82s&42.67	&.9711	&.0675	&33.96	&.9374	&.1074	&34.09	&.9440	&.1284	\\
		20&4.1M&1,251M		&58T&0.68s&42.50	&.9720	&.0664	&33.77	&.9330	&.1134	&33.81	&.9400	&.1244	\\
		12&1.5M &799M		&39T&0.37s&41.85	&.9678	&.0651	&33.65	&.9348	&.1107	&33.16	&.9384	&.1235	\\
		8&0.68M&585M		&27T&0.24s&40.50	&.9645	&.0725	&31.27	&.8982	&.1516	&28.85	&.8815	&.1839	\\
		\hline
		\multirow{2}{*}{\makebox[0.05\textwidth][c]{Methods}}&\multirow{2}{*}{\makebox[0.05\textwidth][c]{\#Param.}}&\multirow{2}{*}{\makebox[0.05\textwidth][c]{Memory}}&\multirow{2}{*}{\makebox[0.05\textwidth][c]{FLOPS}}&\multirow{2}{*}{\makebox[0.05\textwidth][c]{Time}}&\multicolumn{3}{c|}{ICVL}&\multicolumn{3}{c|}{Houston 2018 HSI}&\multicolumn{3}{c}{Pavia City Center HSI}\\
		&&&&&\makebox[0.045\textwidth][c]{PSNR$\uparrow$}&\makebox[0.045\textwidth][c]{SSIM$\uparrow$}&\makebox[0.045\textwidth][c]{SAM$\downarrow$}&\makebox[0.045\textwidth][c]{PSNR$\uparrow$}&\makebox[0.045\textwidth][c]{SSIM$\uparrow$}&\makebox[0.045\textwidth][c]{SAM$\downarrow$}&\makebox[0.045\textwidth][c]{PSNR$\uparrow$}&\makebox[0.045\textwidth][c]{SSIM$\uparrow$}&\makebox[0.045\textwidth][c]{SAM$\downarrow$}\\
		\hline
		TRQ3D & 0.68M & 675M	&1.8T&0.58s& 39.73 & .9491 & .0869	& 32.55 & .9194 & .1241	&28.23	&.8665	&.1961\\
		SST & 4.1M & 1,600M 	&0.7T&1.58s&39.22 	&.9626	&.0743 	& 31.07 & .9166 & .1390 &31.87	&.9234	&.1676\\
		SERT&1.9M&766M&1.7T&0.29s&39.13&.9679&.0963&31.31&.9457&.1517&32.16&.9544&.1384\\
		\hline
		\end{tabular}}
	 \end{table*}

\subsubsection{Earth Observing-1 HSI}
The quantitative results for Earth Observing-1 HSI are shown in Table~\ref{tab:eo1}, where SSUMamba is shown to be the best-performing method among all those evaluated. The visualization results on the Earth Observing-1 HSI, as depicted in Fig.~\ref{fig:eo1_noisy}, present significant challenges due to heavier impulse and stripe noise. While E-3DTV effectively addresses the strips and impulse noise, it results in blurred edges. Among the deep learning-based methods, NSSNN gets a color distortion.
T3SC and TRQ3D successfully mitigate most of the strips but suffer from blurring due to inadequate long-range spatial-spectral correlation modeling. MAC-Net, SST, and SERT struggle to eliminate all the strips, with the SST introducing color distortion. In contrast, SSUMamba effectively leverages spatial and spectral information to yield sharper edges and more accurate colors. The spectral reflectances in Fig.~\ref{fig:eo1_pixel} reveal jitters and discontinuities in the noisy HSI. Notably, SSUMamba achieves smoother and more continuous spectral reflectance.

\subsection{Ablation Study}

Under the case of mixture noise, we present the performance of SSUMamba on different modules to demonstrate their individual contributions to performance. Additionally, we evaluate the impact of network width on the performance of SSUMamba.

\subsubsection{Effectiveness of Scan Schemes and Components}
We evaluate the efficacy of the scan schemes and components introduced in SSUMamba. Specifically, we evaluate the impact of the spatial-spectral continuous scan (SSCS), bidirectional state space model (SSM), and residual block. The results are presented in Table~\ref{tab:ab_module}. The performance with SSCS scan schemes is presented in the first row, while that with a single sweep scan scheme is presented in the second row. SSCS scan uses six continuity scan schemes, enabling the SSM to leverage more effective long-range spatial-spectral correlations. Therefore, better performance is achieved in most cases.    Additionally, the bidirectional SSM enables scanning the HSI in both forward and backward directions, further enhancing the spatial-spectral long-range dependencies modeling. The residual block helps model local spatial-spectral correlations and enhances texture preservation. Consequently, the network equipped with SSCS, bidirectional SSM, and residual block achieves optimal performance.

\subsubsection{Impact of Width}

We conducted experiments to explore the impact of network width on the performance of SSUMamba. Varying the number of channels is crucial in determining network width. In our setup, the 6 SSCS Mamba blocks are configured with channel numbers set as [$C, C \times 2, C \times 2, C \times 4, C \times 4, C \times 8$], where $C$ represents the base channels. We compared the performance of SSUMamba across different numbers of base channels, specifically 32, 24, 20, 12, and 8. Additionally, we compared the results with transformer-based methods, TRQ3D, SST, and SERT. The outcomes are summarized in Table~\ref{tab:ab_width}, which includes details on the number of parameters, memory consumption per batch during training, and floating-point operations per second (FLOPS) and inference time in ICVL testing set for networks of varying widths.

Our findings confirm that as the network width of SSUMamba increases, its performance improves. Compared with transformer-based methods, which suffer from the $\text{O}(n^2)$ complexity, our SSUMamba achieves better performance in the same number of parameters or memory consumption. Specifically, SSUMamba outperforms TRQ3D and SST by providing superior denoising in 8-channel and 20-channel settings, respectively, while using a similar number of parameters and less memory. Compared to SERT, SSUMamba with 12 channels uses a similar number of parameters and memory but achieves higher denoising performance. Additionally, thanks to the parallel optimization of Mamba, our SSUMamba achieves higher throughput indicated by larger FLOPS and faster inference compared to transformer-based methods with a similar number of  parameters.

\section{Conclusion}\label{sec:conclusion}


This paper introduces SSUMamba, a novel approach designed to address the challenge of modeling long-range spatial-spectral correlations in HSI denoising. SSUMamba is based on the spatial-spectral continuous scan (SSCS) Mamba block, which consists of a residual block for modeling local spatial-spectral correlations and a bidirectional state space model (SSM) with the SSCS scheme to exploit long-range spatial-spectral dependencies and continuity in various directions. By simultaneously modeling both local and global spatial-spectral correlations, our proposed method achieves state-of-the-art performance. Furthermore, due to the effective long-range dependencies handled by Mamba, our method occupies fewer parameters, consumes less memory, and has faster inference time.  In our future work, we plan to explore more efficient and robust network designs to achieve faster and higher-performance HSI denoising. Additionally, we aim to integrate our SSCS Mamba block into other tasks, such as HSI superresolution and classification, where leveraging long-range spatial-spectral correlations is crucial for enhancing the effectiveness of these applications.

\appendices
\bibliographystyle{IEEEtran}
\bibliography{IEEEabrv,umamba}
		
\end{document}